\documentclass[twocolumn,superscriptaddress,showpacs,aps,pre]{revtex4-2}

\usepackage[english]{babel}
\usepackage[dvips]{graphics}
\usepackage{graphicx,epsfig}
\usepackage{amsmath}
\usepackage{amsfonts}
\usepackage{amssymb}
\usepackage{xcolor}
\usepackage{multirow}
\usepackage[normalem]{ulem}
\usepackage{booktabs}
\usepackage{float}
\usepackage{makecell}
\usepackage{array}
\usepackage{url}
\usepackage{breakurl}
\usepackage[breaklinks]{hyperref}
\usepackage{placeins}  
\usepackage{amsthm} 
\usepackage{placeins} 
\usepackage{tabularx} 
\usepackage{array}
\usepackage{bm}
\usepackage{lineno}

\theoremstyle{definition} 
\newtheorem{definition}{Definition}[section]

\begin{document}

\title{%
Ontological differentiation as a measure of semantic accuracy
}

\author{P. García-Cuadrillero}
\email{pablo.garcia.cuadrillero@upm.es}
\affiliation{Complex Systems Group, Department of Applied Mathematics,
   Universidad Polit\'ecnica de Madrid,
   Avenida Juan de Herrera 6, 28040 Madrid, Spain.}

\author{F. Revuelta}
\email{fabio.revuelta@upm.es}
\affiliation{Complex Systems Group, Escuela T\'ecnica Superior de Ingenier\'ia Agron\'omica, Alimentaria y de Biosistemas, Universidad Polit\'ecnica de Madrid, Avenida Puerta de Hierro 2-4, 28040 Madrid, Spain.}

\author{J. A. Capitán}
\email{ja.capitan@upm.es (corresponding author)}
\affiliation{Complex Systems Group, Department of Applied Mathematics,
   Universidad Polit\'ecnica de Madrid,
   Avenida Juan de Herrera 6, 28040 Madrid, Spain.}

\begin{abstract}

Understanding semantic relationships within complex networks derived from lexical resources is fundamental for network science and language modeling. While network embedding methods capture contextual similarity, quantifying semantic distance based directly on explicit definitional structure remains challenging. Accurate measures of semantic similarity allow for navigation on lexical networks based on maximizing semantic similarity in each navigation jump (Semantic Navigation, SN). This work introduces Ontological Differentiation (OD), a formal method for measuring divergence between concepts by analyzing overlap during recursive definition expansion. The methodology is applied to networks extracted from the Simple English Wiktionary, comparing OD scores with other measures of semantic similarity proposed in the literature (cosine similarity based on random-walk network exploration). We find weak correlations between direct pairwise OD scores and cosine similarities across $\sim$~2 million word pairs, sampled from a pool representing over 50\% of the entries in the Wiktionary lexicon. This establishes OD as a largely independent, definition-based semantic metric, whose orthogonality to cosine similarity becomes more pronounced when low-semantic-content terms were removed from the dataset. Additionally, we use cumulative OD scores to evaluate paths generated by vector-based SN and structurally optimal Shortest Paths (SP) across networks. We find SN paths consistently exhibit significantly lower cumulative OD scores than shortest paths, suggesting that SN produces trajectories more coherent with the dictionary's definitional structure, as measured by OD. Ontological Differentiation thus provides a novel, definition-grounded tool for analyzing, validating, and potentially constructing navigation processes in lexical networks.
\end{abstract}

\maketitle

\section{Introduction}
\label{sec.intro}

The structure and dynamics of language represent a fascinating domain for complex systems science \cite{CanchoFerrerSole2001, Resnik1995, BoccalettiEtAl2006}. Lexical resources, such as dictionaries and thesauri, when represented as networks where words are nodes and definitional or synonymy relationships form links, exhibit intricate topologies reflecting the underlying semantic organization \cite{SigmanCecchi2002, MotterEtAl2002, SteyversTenenbaum2005}. Understanding how meaning is encoded in these structures and how information propagates through them is crucial not only for linguistics  \cite{Vitevitch2008}, but also for advancing artificial intelligence, particularly in the development and validation of large language models and knowledge representations \cite{BordesWeston2014, NickelKiela2017}.

A common approach to capture semantics involves embedding network nodes in vector spaces, often based on network connectivity patterns \cite{PerozziAlRfou2014} or distributional statistics \cite{MikolovEtAl2013, PenningtonEtAl2014}. For example,  the Random Inheritance Method (RIM) generates such vectors by simulating random walks or information diffusion over the network \cite{BorgeHolthoeferArenas2010, FoussEtAl2007, NohRieger2004}, so that the vectors encode the probability of visiting each node (starting from every other node) after a fixed number of jumps of the associated Markov chain. Similarity between these feature vectors (e.g., cosine similarity) is then used as a proxy for semantic relatedness. This enables specific navigation strategies, a classic problem in the study of complex networks \cite{AdamicEtAl2001}, such as the local-based semantic navigation approach proposed by Capitán et al.~\cite{LocalBasedSNavPaper}, which utilizes vector similarity to jump to the neighbor most similar in each navigation step, thus using Markovian dynamics on the network to guide paths between concepts. However, these embedding-based similarities and the paths they generate reflect statistical correlations or higher-order network proximity, rather than solely the explicit, declarative content of definitions \cite{HughesRamage2007}. Furthermore, standard graph-theoretical measures such as the shortest path distance often prioritize structural economy over semantic coherence, potentially traversing definitionally unrelated concepts. There remains a need for quantitative methods that directly utilize the intrinsic structure of definitions to measure semantic distance or path coherence, providing a complementary perspective to embedding-based approaches and a tool for evaluating semantic processes on these networks.

In this paper, we introduce and formalize Ontological Differentiation (OD), a method designed to quantify, from first principles, the divergence between concepts using the recursive structure of their definitions within a lexical network. At an intuitive level, OD works by repeatedly expanding each initial concept into the terms that appear in its definition, and then expanding those terms in turn, producing successive layers of related words.
When a term is repeated during this process, a so called \emph{cancellation} occurs, signaling a point of semantic overlap. The framework allows for different variants based on specific cancellation rules. In Strong Ontological Differentiation (SOD), the focus of this work, cancellations are triggered only by overlaps between elements originating from different starting concepts. In contrast, Weak Ontological Differentiation (WOD) cancels an element if it has appeared anywhere before, regardless of its origin. The process continues until a termination condition is met, at which point the final OD score is calculated as a weighted sum reflecting the depth at which cancellations occurred.
We present WOD primarily to illustrate the framework's flexibility~\cite{SM},
(see Sec.~\ref{sec.theory}
for the precise rules and a worked SOD example).

Our primary goals are fourfold: (1) to provide a rigorous theoretical framework for defining OD as a meaningful measure of semantic proximity between terms; (2) to investigate the relationship between the definition-based SOD metric and the network-embedding-based cosine similarity derived from RIM; (3) to assess the robustness of these findings by applying our methods to networks constructed from the Simple English Wiktionary under three different definition processing schemes (ground-filtered, random removal, and targeted removal), simulating variations in corpus quality and authorship; and (4) having established the nature of the relationship between SOD and cosine similarity across these conditions, to apply cumulative SOD scores as an independent benchmark to evaluate the semantic coherence of paths generated by different navigation algorithms. Specifically, we use this framework to compare paths generated by Semantic Navigation (SN) \cite{LocalBasedSNavPaper}, which optimizes RIM  vector-based similarities, and standard Shortest Path (SP) navigation, which minimizes the overall path weight, using a large-scale sample of word pairs. 

Our results demonstrate that, while the precise correlation between SOD and RIM cosine similarity varies with the definition processing scheme, SOD consistently acts as a largely independent measure of definitional content, a conclusion supported by both large-scale quantitative analysis and focused qualitative examples that reveal the distinct semantic logic of each metric. Furthermore, and critically for its utility as an evaluation metric, we find that SN paths consistently exhibit lower cumulative SOD scores than SP paths across all tested network conditions. This suggests that SOD quantitatively validates the intuition that SN generates more definitionally coherent trajectories through the semantic network, a finding robust to significant variations in the underlying definitional structure, and it motivates our proof-of-concept exploration of Ontological Navigation, which shows that SOD can serve as a principled navigation heuristic in its own right.

Several prior methods also exploit dictionary or ontology definitions to measure semantic similarity.
That is the case of, e.\,g.,
Extended Gloss Overlap (EGO) \cite{banerjee2002lesk},
where
two glosses are treated as flat pieces of text,
and similarity is measured by the number and length of overlapping word sequences between them, so similarity is a function of the direct word overlap between glosses.
As a consequence,
EGO is fundamentally a non-recursive, single-step comparison.
Conversely,
simDEF~\cite{pesaranghader2016simdef}
represents, in turn, each definition as a bag of words (possibly extended to neighbors in an ontology) and compares these profiles via co-occurrence statistics.
Therefore, it converts definitions into a vector space representation and then computes similarity using a geometric metric
like cosine similarity.
EGO and simDEF compute, therefore,
similarity as the degree of static overlap between glosses or definition sets. 

Alternatively,
recent advances in 
artificial intelligence,
such as node2vec~\cite{Grover2016node2vec}, Graph Neural Networks, and Large Language Models, provide powerful embedding-based approaches to semantic similarity.
These methods produce high-dimensional vector representations trained on distributional or structural data.

In contrast to the previous approaches,
OD provides an entirely symbolic and definition-grounded
process-based measure with two key features.
First, it is fundamentally recursive and depth-weighted: expansions unfold level by level, and the score explicitly reflects not only how many overlaps occur but also how quickly and how abundantly they appear.
Second, it is governed by explicit termination conditions, halting only when one concept’s definitional structure has been fully accounted for by the other.
In this sense, the OD score captures the cumulative effort required for one concept to semantically reach the other, providing a conceptually distinct, combinatorial notion of semantic divergence compared to overlap-based or vector-space methods.
Third, in contrast to other artificial-intelligence approaches,
OD requires no training and yields an interpretable semantic distance based on recursive definitional overlap.
Thus,
OD distinguishes as a complementary technique 
since
it offers a transparent benchmark for semantic coherence that can inform, evaluate, or be integrated with embedding-based techniques.

The remainder of this paper is structured as follows: Section~\ref{sec.theory} details the theoretical framework of Ontological Differentiation and its variants. Section~\ref{sec:navigation} describes the Semantic Navigation and Shortest Path algorithms used for comparison. Section~\ref{sec:application} outlines the three definition processing schemes leading to distinct datasets, network construction, vector computation, the SOD evaluation framework, the design of our large-scale experiment (including details on lexicon size and sampling), and an illustrative example. Section~\ref{sec:results} presents our main findings: it first provides a quantitative and qualitative comparison between SOD and cosine similarity, then shows the results comparing the cumulative SOD scores of SN and SP paths, and culminates in a proof-of-concept exploration of Ontological Navigation. Finally, Section~\ref{sec:conclusions} discusses the implications of our findings.
In addition, the SM~\cite{SM}
provides conceptual examples illustrating OD computation, detailed distributional plots, an analysis of the method's computational scalability, a study on its sensitivity to polysemy, and a crucial extended analysis that investigates the confounding effects of network topology and path length.

\section{First-principle definition of semantic similarity}
\label{sec.theory}
The fundamental premise of our approach is that concepts within a structured lexicon, such as a dictionary, are constituted by their definitions—compositions of other concepts from the same lexicon. This inherently relational and compositional nature implies that the semantic ``distance'' or divergence between any two concepts, $P_A$ and $P_B$, can be understood as a measure of the definitional transformation required for their underlying structures to align. Specifically, we quantify this divergence by examining the extent and depth of recursive definitional expansion needed before the constituent elements of $P_A$ significantly overlap with those of $P_B$.

We formalize this through a framework termed Ontological Differentiation (OD). The ``ontological'' aspect reflects the view that this method addresses the relatedness of entities (here, lexical concepts) that are interdefined and exist through their compositional relationships. OD posits that, if concept $P_A$ is semantically proximal to $P_B$, their definitions, when iteratively expanded, will reveal shared constituent concepts at relatively shallow levels of recursion. Conversely, greater semantic distance implies that such convergence upon common definitional components will only manifest, if at all, after more extensive recursive unfolding, indicating a more substantial structural differentiation between them. OD operationalizes this by systematically tracking the repetition of elements—signifying shared definitional components—during this recursive expansion. The resulting divergence score is then a function of the depth at which these critical overlaps, which we name as cancelations, occur. 

While the primary application and empirical validation in this paper focus on lexical semantics, the core OD methodology—quantifying divergence through the recursive depth needed for compositional alignment via specific cancelation and termination rules—is conceived as a generalizable framework. It holds potential for analyzing relationships in any system where entities are characterized by their interconnections and constituent parts, with the particular cancelation and termination rules adaptable to the specific nature of the entities and their relationships in the domain of interest.

The formal development of OD begins by representing a lexicon as a set of interdefined concepts:

\begin{definition} 
    Let $\mathbf{U}$ be a set of elements $\{P_{1},...,P_{n}\}$ of size $n\in\mathbb{N}$, for which every element $P_{i}\in\mathbf{U}$ is itself a set of size $m_i=|P_i|$ (can vary per element) of elements in $\mathbf{U}$, that is $P_{i}=\{P_{i_{1}},...,P_{i_{m_i}}\}$ where every $P_{i_{j}}\in\mathbf{U}$. This represents a collection of concepts where each concept is defined by a set of other concepts in the collection.
\end{definition}

\subsection{Ontological differentiation}
We first define the process of expanding definitions recursively.
\begin{definition} 
    We call $R$ the \emph{read function}, an operation such that for any $P_{i}\in\mathbf{U}$, $R(P_{i})$ yields the set $\{P_{i_{1}},...,P_{i_{m_i}}\}$ of elements in $\mathbf{U}$ constituting its definition. $R^{n}(P_{i})$ denotes the multiset of elements obtained by applying the read function recursively $n$ times.
\end{definition}
Intuitively, this function ``reads'' the definition of term $P_i$ in the dictionary and outputs the set of words appearing in the definition. Formally, if we have as input some element $P_{i}\in\mathbf{U}$ whose content is given as $P_{i}= \{P_{i_{1}},...,P_{i_{m}}\}$, the read function $R^{1}(P_{i})$ will produce the multiset containing $P_{i_{1}},...,P_{i_{m}}$. The parameter $n$ specifies how many times this expansion operation is carried out, being $n=0$ just the reading $R^{0}(P_i)=\{P_i\}$. For instance, $R^{2}(P_{i})$ represents the multiset formed by applying $R$ to each element in $R^1(P_i)$. This can be generalized as $R^{n}(P_{i}) = \bigcup_{P_k \in R^{n-1}(P_i)} R(P_k)$.

The read function can be applied not only to a single term in the dictionary, but to a subset of words. Then, the read function of the subset is naturally defined as the union of terms appearing in any definition of each element of the subset (i.e, the union of the output of the read function applied to each individual term in the subset).
\begin{definition} 
    Let $\mathcal{A} = \{P_1, ..., P_k\} \subseteq \mathbf{U}$ be an initial set of $k$ elements. Then, the read function $R^{\omega}(\mathcal{A})$ is defined  as the collection of all elements generated by expanding each $P_s \in \mathcal{A}$ up to $\omega$ levels using the read function $R$. That is, $R^{\omega}(\mathcal{A}) = \bigcup_{n=0}^{\omega} \bigcup_{s=1}^{k} R^n(P_s)$.
\end{definition}
The core idea of our definition of semantic proximity is related to the repetition of terms during this expansion process applied to the initial set $\mathcal{A}$. This is governed by specific rules. As discussed in the next subsection, different cancelation and termination rules produce different measures of semantic proximity. 
\begin{definition} 
    A \emph{cancelation rule} $\kappa$ determines the conditions under which an instance of an element $\alpha$ appearing in the expansion $R^n(P_s)$ (originating from $P_s \in \mathcal{A}$ at level $n$) is marked as ``canceled'' or ``repeated'' based on its appearances elsewhere within the overall expansion $R^{\omega}(\mathcal{A})$.
\end{definition}
Cancelation occurs when words arise repeatedly during the recursive expansion $R^{\omega}(\mathcal{A})$. As this expansion defines a hierarchy of levels (depending on the recursion level $n$), words will be more distant if repetitions arise higher in this hierarchy. 

This definition is intentionally stated in general terms, abstracting away from the specifics of which expansions are compared. To see how such rules operate on expansions and how overlaps trigger cancellations under SOD and WOD, we refer the interested
reader to the worked examples in Sec.~\ref{subsubsec:sod}
and in the SM~\cite{SM}.
\begin{definition} 
    A \emph{termination rule} $\tau$ determines the maximum level of expansion $\omega$, based on the state of canceled/uncanceled elements resulting from applying the cancelation rule $\kappa$.
\end{definition}
    The process of applying a specific cancelation rule $\kappa$ and termination rule $\tau$ to an initial set $\mathcal{A}$ is denoted $F_{\kappa\tau}(\mathcal{A})$. This process determines which elements get canceled at each level $n$ (from $0$ to $\omega$) and the final expansion depth $\omega$. As a result, $F_{\kappa\tau}(\mathcal{A})$ yields a finite, recursive expansion of the terms initially included in $\mathcal{A}$. We name the process $F_{\kappa\tau}(\mathcal{A})$ as the \emph{Ontological Differentiation} of set $\mathcal{A}$.
    
    Let $C_{nm}$ be the count of element instances originating from $P_m \in \mathcal{A}$ that are marked as canceled specifically at level $n$ according to rule $\kappa$.
The above definitions allow us to quantify the divergence of this recursive definition expansion. 
\begin{definition}
    The \emph{score} of the Ontological Differentiation process $F_{\kappa\tau}$ applied to $\mathcal{A}$ is given by the weighted sum of the number of canceled elements at each level:
    \[
        \text{Score}(F_{\kappa\tau}(\mathcal{A})) = \sum_{n=0}^{\omega} n \sum_{m=1}^{k} C_{nm}.
    \]
\end{definition}

This score quantifies the total ``differentiation effort.'' By weighting cancelations by the recursion level $n$ at which they occur, the score directly reflects our foundational premise: concepts are considered more distant (i.e., require a greater extent of definitional unfolding to find common ground) if their shared constituent elements—leading to cancelations—only emerge at deeper levels of their recursive expansion. Thus, the scoring method penalizes overlaps that occur at greater depths (higher $n$) and, conversely, gives less weight to (or ``benefits'') those that occur at shallower levels (lower $n$), signifying closer semantic proximity. While this specific weighting scheme (multiplying by $n$) aligns with our focus on definitional depth in lexical networks, the general OD framework allows for this scoring function itself to be modified (e.g., using different weighting functions or incorporating other factors beyond depth) to suit the specific characteristics and desired interpretations of divergence in other domains or scenarios.

We finally define the overall distance between the elements in the set $\mathcal{A}$, which leads to a measure of semantic proximity between the terms in this set.
\begin{definition}
    The \emph{Ontological Differentiation distance} $\text{OD}(\mathcal{A})$ between the elements in the set $\mathcal{A}$ is defined as the score obtained from a specific differentiation process $F_{\kappa\tau}$: $\text{OD}(\mathcal{A}) = \text{Score}(F_{\kappa\tau}(\mathcal{A}))$. 
\end{definition}
Naturally, the specific OD distance value depends on the chosen rules $\kappa$ and $\tau$, as well as the chosen scoring function.

It is important to note that the OD score, particularly the SOD variant used in our analysis, can grow very rapidly. The score is calculated as a sum of products, where one factor is the recursion level, $n$. As definitions are expanded recursively, both the number of potential cancellations and the recursion level $n$ increase, leading to a combinatorial growth in the potential score. Consequently, SOD scores for moderately distant concepts can easily span many orders of magnitude. This characteristic necessitates the use of logarithmic transformations for effective visualization and statistical analysis, as shown in our results.

\subsection{Types of ontological differentiation}
\label{subsec:od_types} 

The general OD framework allows for various types, each defined by specific cancelation and termination rules. Here we introduce Strong Ontological Differentiation (SOD)
first, due to its intuitive alignment with how semantic divergence might be perceived based on definitional content. The primary distinction between the variants we discuss lies in what constitutes a meaningful repetition. Consider the expansion originating from two initial concepts, $P_A$ and $P_B$. 
SOD
employs a strict, cross-branch rule: an element is canceled only if it appears in the expansion of $P_A$ and also appears in the expansion of $P_B$. This directly measures the convergence of distinct definitional paths. 
For that reason, in this work we focus primarily on SOD.

However, to illustrate the flexibility of our framework, 
besides SOD,
we also briefly describe a second type, Weak Ontological Differentiation (WOD), which uses a more general rule: an element is canceled if it has appeared anywhere before in the total expansion, regardless of which branch it originated from. Finally, we introduce Great Ontological Differentiation (GOD), which is key in bounding the number of recursion steps in the expansion process. Conceptual examples further illustrating the computation of WOD and GOD scores, based on abstract concepts, are provided in the 
SM~\cite{SM}.

\subsubsection{Strong ontological differentiation (SOD)}
\label{subsubsec:sod}

Strong Ontological Differentiation (SOD) quantifies divergence by identifying elements that appear in expansions originating from different initial concepts. Repetitions occurring solely within the expansion of a single initial concept do not cause cancelation unless the element also appears in the expansion of another initial concept from the starting set $\mathcal{A}$. We formalize SOD cancelation and termination rules as follows:

\begin{definition} 
    ($\text{SOD}$ cancelation rule, $\kappa_S$)  An instance of element $\alpha$ generated at level $n$ from $P_s \in \mathcal{A}$ is canceled if $\alpha$ also appears in the expansion generated from a different initial element $P_r \in \mathcal{A}$ (where $r \neq s$) at any level $h \le n$. Formally, if $\alpha \in R^n(P_s)$ and there exists $\beta \in R^h(P_r)$ such that $\alpha=\beta$ where $r \neq s$ and $h \le n$, then $\alpha$ is canceled.
\end{definition}

\begin{definition} 
    ($\text{SOD}$ termination rule, $\tau_S$) The expansion stops at level $\omega$ if, after applying the cancelation rule $\kappa_S$ up to level $\omega$, all elements generated from at least one of the initial sets $P_s \in \mathcal{A}$ at level $\omega$ (i.e., all elements in the multiset $R^\omega(P_s)$) have been canceled.
\end{definition}

The specific design of these SOD rules is motivated by our aim to measure the definitional ``distance'' between two concepts $P_s$ and $P_r$ by tracking how one concept's expanded definition progressively incorporates elements from the other. The $\kappa_S$ cancelation rule focuses exclusively on \emph{cross-side} repetitions: an element is only canceled if it originates from one initial concept (e.g., $P_s$) and is also found within the expansion of the other initial concept ($P_r$). This distinction is crucial because we are interested in the convergence of initially distinct definitional paths; internal repetitions within a single concept's expansion do not, by themselves, signify progress towards alignment with another concept. 
The $\tau_S$ termination rule is triggered when all elements comprising a specific expansion level of one initial concept (e.g., all words in $R^1(P_s)$) are found to be ``covered'' or canceled by elements from the expansion of the other initial concept(s). This signifies a point where, at that particular depth of expansion, one concept's definitional layer has been fully accounted for within the broader definitional structure of the other, indicating a significant degree of definitional subsumption or overlap.

We illustrate how SOD operates with the aforementioned rules by calculating the score for the pair of words `job' and `business'. 
We will demonstrate the SOD calculation assuming that the definitions for these terms and their constituents are as provided in the ``Definitions'' column of Table~\ref{tab:sod_job_business_example}, as might be extracted from a lexical resource after a filtering process similar to that described later in Section~\ref{subsec:data_prep}.

The SOD process, detailed in Table~\ref{tab:sod_job_business_example}, unfolds by iteratively expanding these definitions. After each level of expansion, the SOD cancelation rule ($\kappa_S$) is applied by looking for elements appearing on both the `job' and `business' sides across all currently expanded levels. Subsequently, the SOD termination rule ($\tau_S$) is checked. In Table~\ref{tab:sod_job_business_example}, elements that are ultimately canceled by this process are shown in bold.

\begin{table*}[t]
  \centering
  \small
  \setlength\tabcolsep{4pt}
  \renewcommand{\arraystretch}{1.3}
  \begin{tabular}{
      |>{\raggedright\arraybackslash}m{0.24\textwidth}
      |>{\centering\arraybackslash}m{0.18\textwidth}
      |>{\raggedright\arraybackslash}m{0.13\textwidth}
      |>{\raggedright\arraybackslash}m{0.13\textwidth}
      |>{\centering\arraybackslash}m{0.10\textwidth}|
    }
    \hline
    \makecell[c]{\textbf{Definitions}}
      & \makecell[c]{\textbf{Level}}
      & \makecell[c]{\(\bm{P_s}=\mathrm{job}\)}
      & \makecell[c]{\(\bm{P_r}=\mathrm{business}\)}
      & \makecell[c]{\textbf{Scores}} \\
    \hline

    \multirow[c]{3}{0.24\textwidth}{
      $\!$job: [pay, money, service]\\[9pt]
      business: [buy, sell, service]\\[9pt]
      pay: [give, money]\\[9pt]
      money: [coin, note]\\[9pt]
      service: [work, consume]\\[9pt]
      buy: [pay, money, obtain]\\[9pt]
      sell: [give, gain, money]
    }
    & \makecell[c]{Level 0\\\(R^0(\mathrm{job,business})\)}
    & \makecell[c]{job}
    & \makecell[c]{business}
    & \makecell[c]{\(0\times0=0\)} \\
    \cline{2-5}

    & \makecell[c]{Level 1\\\(R^1(\mathrm{job,business})\)}
    & \makecell[c]{\textbf{pay}\\\textbf{money}\\\textbf{service}}
    & \makecell[c]{buy\\sell\\\textbf{service}}
    & \makecell[c]{\(4\times1=4\)} \\
    \cline{2-5}

    & \makecell[c]{Level 2\\\(R^2(\mathrm{job,business})\)}
    & \makecell[c]{\textbf{give}\\\textbf{money}\\coin\\note\\\textbf{work}\\\textbf{consume}}
    & \makecell[c]{\textbf{pay}\\\textbf{money}\\obtain\\\textbf{give}\\gain\\\textbf{money}\\\textbf{work}\\\textbf{consume}}
    & \makecell[c]{\(10\times2=20\)} \\
    \hline

    \multicolumn{5}{|c|}{\textbf{Total distance = 24}} \\
    \hline
  \end{tabular}
  \caption{Example showing the calculation of the score SOD(job, business). The definitions of $P_s=$ `job' and $P_r=$ `business' include the terms `pay', `money', `service', `buy' and `sell', which are expanded recursively according to their definitions (first column). The different multisets $R^n(P_i)$, $i\in\{s,r\}$, obtained by recursive expansion at the $n$-th level of recursion are shown in columns $P_s$ and $P_r$, together with each level's score (shown in the last column). Canceled elements are marked in boldface. In this example, after the second level expansion we observe that all the terms in the `job' column at level 1 have appeared at levels less or equal than 2, so the process terminates at the second level. The score is calculated as the overall number of repetitions weighted by the level of recursion at which those repetitions occurred.}
  \label{tab:sod_job_business_example}
\end{table*}

The step-by-step application of this SOD process to the (`job', `business') pair unfolds as follows:

\begin{itemize}
    \setlength{\leftmargini}{0.5cm}
    \setlength{\leftmarginii}{0.5cm}
    \item \textbf{Level 0 ($R^0$):}
        \begin{itemize}
            \item[$\circ$] \textit{Expansion:} `job' and 'business' are the initial elements.
            \item[$\circ$] \textit{Cancelations:} None, as no prior elements or cross-side repetitions exist.
            \item[$\circ$] \textit{Termination check:} No side is fully canceled at any level. The process continues.
        \end{itemize}

    \item \textbf{Level 1 ($R^1$):}
        \begin{itemize}
            \item[$\circ$] \textit{Expansion:} `job' yields \{pay, money, service\}; 'business' yields \{buy, sell, service\}.
            \item[$\circ$] \textit{Cancelations:} Considering levels L0-L1, we observe that the element `service' appears in both $R^1(\text{job})$ and $R^1(\text{business})$. Thus, `service' is marked for cancelation (in boldface) on both sides at level L1.
            \item[$\circ$] \textit{Termination Check:} No side at any level is fully canceled yet. The process continues to the next level of expansion.
        \end{itemize}

    \item \textbf{Level 2 ($R^2$):}
        \begin{itemize}
            \item[$\circ$] \textit{Expansion:} Elements from $R^1(\text{job})$ and $R^1(\text{business})$ are expanded according to the corresponding definitions (see Table~\ref{tab:sod_job_business_example}).
            \item[$\circ$] \textit{Cancelations:} Considering levels L0-L2, we recognize the following repetitions: (i) `give', `money', `work', `consume' appear in both $R^2(\text{job})$ and $R^2(\text{business})$; (ii) `pay', from $R^2(\text{business})$, also appears in $R^1(\text{job})$; (iii) `money', from $R^2(\text{business})$, also appears in $R^1(\text{job})$.
            \item[$\circ$] \textit{Termination check:} After all cancelations are resolved, all elements of the multiset $R^1(\text{job}) = \{\text{pay, money, service}\}$ are now marked as canceled. Therefore, the termination condition $\tau_S$ is met. The process stops, and the final score is calculated as shown in the table.
        \end{itemize}
\end{itemize}

This example illustrates that cancelations in SOD are determined by cross-side repetitions considering all generated elements up to the current expansion depth. The process terminates when all elements originating from one of the initial concepts at a specific level of its expansion are canceled.

\subsubsection{Weak ontological differentiation (WOD)}
\label{subsubsec:wod}

For completeness, we note a less restrictive variant, Weak Ontological Differentiation (WOD). Unlike SOD, which cancels elements only when they appear in expansions of different initial concepts (cross-branch repetitions), WOD cancels any repeated element regardless of its origin branch or level. This variant illustrates that OD is a general framework with multiple possible cancellation rules. 

A full definition and a detailed, step-by-step worked example of WOD (including its termination and scoring) are provided in the SM~\cite{SM}.
This appendix example, like others provided for conceptual clarity, presents a more exhaustive, visual walkthrough of the iterative expansion and cancelation process. Such detailed tracking is excellent for understanding the core logic but is not the computationally efficient method used for large-scale application. The actual calculations performed in this study utilize an optimized algorithm that tracks sets of uncanceled and canceled elements at each level. The implementation of this algorithm is available in our repository~\cite{ODCodeRepo}, and a detailed analysis of its computational cost, scalability and termination guarantees is presented in the 
SM~\cite{SM}.

\subsubsection{Great ontological differentiation (GOD)}
\label{subsubsec:god}

The Great Ontological Differentiation (GOD) process is not used for scoring divergence itself, but rather to determine a maximum relevant expansion depth, $\omega_{GOD}$. This is crucial for ensuring the finiteness of SOD and WOD calculations, especially in networks with cyclic definitions. GOD employs a distinct termination rule ($\tau_G$) based on the emergence of new unique elements during the expansion, which serve to determine an upper bound for the expansion depth.

\begin{definition} 
    ($\text{GOD}$ termination rule, $\tau_G$) The expansion terminates at level $\omega$ if the set of all elements generated at level $\omega$, $\bigcup_{s=1}^k R^\omega(P_s)$, contains no element $\alpha$ that was not already present in the union of elements from all preceding levels, $\bigcup_{n_p=0}^{\omega-1} \bigcup_{s=1}^k R^{n_p}(P_s)$. We denote this termination level as $\omega_{GOD}$.
\end{definition}
The level $\omega_{GOD}$, illustrated with an abstract example in the SM~\cite{SM},
signifies the point beyond which no new concepts are introduced into the collective definitional expansion. Therefore, it serves as a practical upper bound for SOD/WOD: if the primary termination rule has not been met by $\omega_{GOD}+1$, the OD process can be halted. This prevents infinite loops and ensures the differentiation process has explored all potentially new definitional content before terminating due to stability rather than complete cancelation on one side.

\subsubsection{Other types}
\label{subsubsec:other_types}
The specific definitions chosen for the cancelation rule ($\kappa$) and the termination rule ($\tau$) dictate the behavior of any particular Ontological Differentiation process. The SOD, WOD, and GOD variants presented utilize straightforward rules based on element repetition, origin, and novelty. However, the general OD framework permits the formulation of numerous other rule sets. One could construct variants by modifying the existing conditions, combining elements of different rule types, introducing probabilistic elements, or defining entirely novel criteria for cancelation and termination based on different structural or sequential properties of the expansion. A full exploration of the mathematical possibilities afforded by altering these rules is beyond the scope of this work, where our aim is to introduce the core OD concept and demonstrate the utility of the SOD variant through empirical application.

\section{Navigation methods}
\label{sec:navigation}

In this section, we describe two distinct strategies for finding a route between a source node \(s\) and a target node \(t\) in a semantic network: (1) \emph{semantic navigation} (SN), which uses high‐order connectivity via vector embeddings, and (2) \emph{shortest‐path} (SP) navigation, which finds cost‐minimal routes in the graph.

\subsection{Graph representation}

We model the semantic network as an undirected, weighted graph $G = (V, E)$, where: (i) \(V = \{1, 2, \dots, N\}\) is the set of \(N\) nodes, each corresponding to a processed dictionary entry; and (ii) \(E \subseteq V \times V\) is the set of edges. 

The graph is constructed from the initial directed definition links. First, for each headword $i$ defined by a set of $m_i$ tokens, a directed edge is drawn from $i$ to each token $j$, with weight $w_{i \to j} = 1/m_i$. The network is then symmetrized by ensuring that for every edge $i \to j$ there is also a reverse edge $j \to i$. If the reverse edge $j \to i$ is absent, we add it with the same weight $w_{j \to i} := w_{i \to j}$. If both $i \to j$ and $j \to i$ already exist in the dictionary, we keep both with their original (possibly different) weights, reflecting the distinct outdegrees of $i$ and $j$. This choice is crucial for modeling the real, navigable space of a digital lexicon. For instance, if entry B is in the definition of A, a directed link $A \to B$ exists. However, A may not appear in the definition of B, meaning the reverse link $B \to A$ is absent. A user navigating from A to B can, however, always return to A (e.g., via a ``back'' button). To capture this fundamental bidirectional navigability, we enforce symmetry. This process reflects the principle that a definitional connection, even if unidirectional, implies a mutual semantic association that can be traversed in both directions. While this symmetric network forms the basis of our main results, we conduct a parallel analysis on the original (non-symmetrized) directed network in the 
SM~\cite{SM}
to validate our findings under stricter assumptions.

\subsection{Semantic navigation}
\label{subsec:navigation_sn}
Semantic Navigation provides a method for finding paths between concepts by leveraging the relationships encoded in high-dimensional vector embeddings derived from the network structure. Specifically, we adapt the local-based semantic navigation strategy proposed by Capitán et al.~\cite{LocalBasedSNavPaper}, which operates as a greedy, cosine-similarity-driven walk using vectors generated via exploring network structure using random walks.

\paragraph*{Semantic embeddings.}

The Random Inheritance Method (RIM) \cite{BorgeHolthoeferArenas2010} is a method inspired by exploring the network through random walks. A discrete-time Markov chain can be associated to the random walkers, whose transition probabilities are encoded in a row-stochastic transition matrix $P \in \mathbb{R}^{N\times N}$. In our implementation, this matrix reflects uniform transition probabilities based on node degrees in the symmetrized graph. Denoting by $\text{deg}(i)$ the degree of node $i$ in the symmetrized graph $G$, then the transition matrix is: $P_{ij} = 1 / \text{deg}(i)$ if $(i, j) \in E$ and $i \neq j$, and $P_{ij}=0$ otherwise. In such definition, we ignore self-loops. By construction, it is ensured that $\sum_j P_{ij} = 1$. Note that the weights $w_{ij}$ are not used to construct $P$.

The elements of matrix powers $P^k$ encode the probabilities of jumping between pairs of nodes up to $k$ steps. This motivates the definition, for each node \(i\in V\), of its \emph{semantic vector} $\mathbf{v}_i \in\mathbb R^N$ as a vector encoding the random walk exploration of the network up to a certain inheritance depth level:
\[
  \mathbf v_i \;=\; \sum_{k=1}^{T}\mathbf e_i P^k,
\]
where (i) \(\mathbf e_i\) is the \(i\)th standard basis vector in \(\mathbb R^N\); (ii) \(P^k\) is the \(k\)-step transition probability matrix derived from the Markov chain defined by the stochastic matrix $P$; (iii) \(T\) is the \emph{inheritance depth}, usually chosen to exceed the network's diameter (we use \(T=10\)).

Thus \(\mathbf v_i\) captures the cumulative reachability profile of node \(i\) up to \(T\) hops in the random network exploration defined by $P$. Pairs of semantic vectors $\mathbf v_i$ and $\mathbf v_j$ are then used to measure the cosine similarity between nodes $i$ and $j$.

\paragraph*{Cosine-based network navigation.}
Given a source \(s\in V\) and target \(t\in V\) nodes in the network, SN constructs a path \(\mathcal P = (p_0, p_1, \dots, p_L)\) by iterating:
\begin{enumerate}
  \item \textbf{Initialization:} Set $p_0 = s$, step count $k=0$. Define a maximum path length $L_{\max}$ (it can be chosen as function of the graph diameter).
  \item \textbf{Iteration:} While $p_k \neq t$ and $k < L_{\max}$:
    \begin{enumerate}
      \item Set current node $c = p_k$.
      \item Let $\mathcal{N}(c)$ be the set of unvisited neighbors: $\mathcal N(c) = \{j \mid (c,j)\in E \text{ and } j \notin \{p_0, \dots, p_k\}\}$.
      \item If $\mathcal{N}(c)$ is empty, terminate (failure).
      \item For each $j\in\mathcal N(c)$, compute the cosine similarity between $\mathbf{v}_j$ and $\mathbf{v}_t$:
        \[ \sigma_j \;=\; \frac{\mathbf v_j \cdot \mathbf v_t}{\|\mathbf v_j\| \|\mathbf v_t\|} \]
        (handle potential zero vectors).
      \item Choose the next node in the path as the one that maximizes cosine similarity with the current node $c$: $p_{k+1} = \arg\max_{j\in\mathcal N(c)} \sigma_j$.
      \item Increment $k \leftarrow k+1$.
    \end{enumerate}
  \item \textbf{Termination:} The algorithm stops when $p_k = t$ (success) or $k=L_{\max}$ or $\mathcal{N}(c)$ was empty. It returns a navigation path $\mathcal P = (p_0, \dots, p_k)$.
\end{enumerate}

Because each move selects the neighbor whose embedding is most aligned with the target, SN aims to produce semantically coherent routes, albeit without guaranteeing minimal graph-theoretic cost.

\subsection{Shortest‐path navigation}
As a standard graph-theoretic baseline for comparison with SN, we consider shortest-path navigation. The objective of SP navigation is to find a path \(\mathcal Q^*\) between a given source node \(s\) and target node \(t\) that minimizes the total cumulative edge weight. For any path \(\mathcal Q=(q_0=s, q_1, \dots, q_L=t)\) consisting of \(L\) steps, this cumulative cost is defined as:
\[
  \mathrm{Cost}(\mathcal Q) \;=\; \sum_{\ell=1}^L w_{q_{\ell-1}, q_\ell}
\]
where \(w_{ij}\) are the non-negative, symmetric edge weights defined in Sec.~\ref{sec:navigation}A. 

It is important to note that the term ``shortest path'' in our study refers to this minimum-cost path, which is not necessarily the path with the fewest number of edges (hops). To compute this minimum-cost path \(\mathcal Q^*\), we employ Dijkstra's algorithm \cite{Dijkstra1959}. This widely used algorithm is guaranteed to find the path with the lowest cumulative cost from a single source node to all other nodes in a graph with non-negative edge weights. It operates by iteratively exploring the network outward from the source \(s\), maintaining the lowest cost found so far to reach each node.

\section{Evaluating Navigation Methods with SOD}
\label{sec:application}

To assess the utility of OD as a measure of semantic accuracy and to compare the pathways generated by Semantic Navigation and Shortest Path algorithms, we implemented a computational framework using data derived from the Simple English Wiktionary. This section details the data preparation, network construction, vector computation, evaluation methodology, and the design of the large-scale comparison experiment. Furthermore, an illustrative example is provided at the end of this section (Section~\ref{subsec:examples}) to demonstrate how cumulative SOD scores are used to distinguish between different navigation paths.

\subsection{Data acquisition and definition processing}
\label{subsec:data_prep}

Our primary data source was the Simple English Wiktionary (queried March 2025), chosen for its relatively controlled vocabulary and simpler definition structures. After an initial extraction and indexing process (detailed further in Sec.~\ref{subsec:sampling}), we established a working lexicon of 19,283 unique entries for which definitions were available and which formed the basis for our subsequent experiments.

To assess the robustness of our findings and simulate variations in corpus characteristics (e.g., differing authorship, curation levels, or data sparsity) without introducing the many uncontrolled variables inherent in using entirely different external corpora, we generated three distinct definition datasets from this common 19,283-entry lexicon. This approach of systematically modifying a single source allows for a more controlled and theoretically grounded comparison of how variations in definitional structure impact the metrics. The generation of these datasets involved a foundational ``ground filtering'' stage applied to the definitions of these entries, followed by specific modifications to create each version.

\paragraph*{Ground filtering (common to all datasets).}
Initially, the definitions corresponding to our working lexicon underwent a common set of rigorous filtering and cleaning processes aimed at isolating core semantic relationships and standardizing the lexical base. For entries exhibiting polysemy, we adopted the simplifying convention of using only the first definition provided in the Wiktionary entry. This choice ensures a consistent and reproducible framework for this foundational study. To quantify the impact of this simplification, we conducted a targeted sensitivity analysis (see SM~\cite{SM},
which confirmed that the OD metric is highly sensitive to the specific word sense used. The results show that different senses lead to large, semantically coherent shifts in SOD scores, validating the potential for future work where OD could be extended to handle multiple word senses, for instance by averaging scores or using contextual disambiguation. 

Entries' definitions were then filtered based on their likely part-of-speech (only nouns, verbs, and adjectives retained) to focus on core lexical concepts. Furthermore, heuristics were applied to filter terms potentially representing inflected forms and to identify lemmas for definition tokens, aiming to build the network primarily from base forms. The text was then tokenized into constituent words. These tokens were further filtered: very short or numerical tokens were removed. Finally, to prevent potential trivial results in the subsequent OD calculations, any instance of a headword appearing within its own processed definition was removed.

\paragraph*{Dataset 1: ground-filtered.}
This dataset directly utilizes the definitions resulting from the ``ground filtering'' stage described above, with no further lexical modification of the definition content. It represents the most comprehensive but least specifically curated version of the definitions for the established lexicon and serves as the input for the other two processing schemes.

\paragraph*{Dataset 2: random removal.}
This dataset was created by taking the ``ground-filtered'' definitions and applying a structural perturbation. The ground-filtered definitions for the entries in our lexicon exhibit an average of 4.79 tokens per definition, with 2.32\% of these definitions containing only a single token. For every definition in this set that contained more than one token (thus affecting 97.68\% of the definitions), one token was randomly selected and removed. This process simulates the effect of data sparsity or less comprehensive definitions that might arise from different authorship or corpus limitations.

\paragraph*{Dataset 3: targeted removal.}
Starting from the ``ground-filtered'' definitions, this dataset underwent an additional, specific normalization step. A predefined list of 20 common, high-frequency, and often low-semantic-content words (analogous to stop-word removal) was applied to eliminate these terms from all definitions. Notably, all 20 of these removed words ranked within the top 100 most frequent tokens appearing in the ground-filtered definitions, and 13 of them were within the top 20. This aggressive, targeted filtering, detailed in our online repository~\cite{ODCodeRepo} (which includes the list of removed words), was designed to create a more semantically focused network by removing potentially noisy functional or overly general terms.

These three distinct datasets, each based on the same lexicon but with varying definition content resulting from the ground-filtered, random removal, and targeted removal processing schemes, were then used independently for subsequent network construction, vector computation, and all comparative analyses. 

\subsection{Network construction}
\label{subsec:network}

For each of the three definition datasets (ground-filtered, random removal, and targeted removal), we independently constructed an undirected, weighted graph $G=(V, E)$ to represent the semantic relationships inherent in that specific dataset. The procedure for defining nodes, creating and symmetrizing edges, and assigning edge weights $w_{ij}$ followed the methodology detailed in Sec.~\ref{sec:navigation}A. This crucial step resulted in three distinct network structures, providing the foundation for all subsequent pathfinding analyses.

\subsection{Semantic vector computation}
\label{subsec:vectors}

To enable the semantic navigation algorithm, we computed semantic vectors $\mathbf{v}_i$ for each node within each of the three network structures generated in Sec.~\ref{subsec:network}. Vectors were generated using RIM, following the procedure detailed in Sec.~\ref{subsec:navigation_sn}. The specific implementation for each condition used the topology of the corresponding network to define the transition matrix $P$. This yielded three distinct sets of semantic vectors, ensuring that all SN-based analyses were performed using representations derived directly from the appropriate definitional dataset.

\subsection{Direct comparison of SOD and cosine similarity}
\label{subsec:sod_vs_cosine_eval}

Before using SOD to evaluate navigation paths, it is essential to understand its relationship to established vector-based semantic similarity measures. Therefore, a primary step in our analysis is to investigate the correlation between the proposed definition-based metric and a standard metric derived from network embeddings. Specifically, for a given pair of concepts represented by a source node $s \in V$ and a target node $t \in V$, we compute and compare two metrics:
\begin{enumerate}
    \item The direct pairwise Strong Ontological Differentiation score, $\text{SOD}(s, t)$, calculated as described in Sec.~\ref{sec.theory}.
    \item The cosine similarity, $\cos(\mathbf{v}_s, \mathbf{v}_t)$, calculated between the corresponding RIM vectors $\mathbf{v}_s$ and $\mathbf{v}_t$ (generated as described in Sec.~\ref{subsec:vectors}, based on ~\cite{BorgeHolthoeferArenas2010, LocalBasedSNavPaper}).
\end{enumerate}

Comparing these direct measures allows us to investigate whether the definition-based SOD metric captures similar or different information compared to the vector-embedding-based cosine similarity. This comparison is crucial for interpreting the path evaluation results presented in Sec.~\ref{sec:results}. If SOD and cosine similarity were found to be highly correlated, then lower cumulative SOD scores for SN paths might simply reflect that SOD measures what SN optimizes (cosine similarity). Conversely, a lack of correlation would suggest SOD offers an independent validation based on definitional structure. Establishing this independence strengthens the claim that SOD measures a distinct facet of semantic relatedness, and that the observed alignment between SN paths and lower cumulative SOD is non-trivial. The computation of both direct metrics for each pair in our large-scale experiment facilitates this investigation.

\subsection{Path computation and SOD evaluation framework}
\label{subsec:sod_eval}

Having established the context by comparing direct SOD and cosine similarity, our principal evaluation focuses on using cumulative SOD to quantitatively assess the semantic coherence of paths generated by the SN and SP navigation methods (detailed in Sec.~\ref{sec:navigation}). For a given source-target pair $(s, t)$, we compute both the SN path $\mathcal{P}_{\rm SN}$ (via the greedy-cosine walk) and the SP path $\mathcal{Q}_{\rm SP}$ (via Dijkstra's algorithm using weights $w_{ij}$).

To assess the overall semantic divergence of a path from its origin, we calculate a cumulative SOD score. This score reflects the total differentiation effort required to reach each node on the path from the starting node $p_0=s$. Specifically, for a path $\mathcal{P}=(p_0, p_1, \dots, p_L)$, the cumulative SOD score, $\text{SOD}_C(\mathcal{P})$, is calculated as the sum of the pairwise SOD scores between the source node $p_0$ and every subsequent node $p_i$ ($i=1, \dots, L$) along the path:
\[ 
    \text{SOD}_C(\mathcal{P}) = \sum_{i=1}^{L} \text{SOD}(p_0, p_{i})_{\omega_{GOD}+1}
\]

Here, $\omega_{GOD}+1$ is the maximum expansion depth for each pairwise calculation, which is determined beforehand by computing $\text{GOD}(p_0, p_i)$. This limit, indicated by the $\omega_{GOD}+1$ subscript, is imposed on the SOD calculation to guarantee termination and prevent potential infinite loops arising from cyclic definitions within the network. If the calculation reaches level $\omega = \omega_{GOD}+1$ without the termination rule being met, the process is halted at this level. This condition prevents infinite computation by ending that pair operation. 

It is worth noting that alternative definitions for a path's cumulative SOD score exist. For instance, one could sum the SOD scores between each intermediate node $p_i$ ($i=0, \dots, L-1$) and the final target node $p_L=t$. While potentially valid, we selected the current formulation, $\sum_{i=1}^{L} \text{SOD}(p_0, p_{i})$, as it directly measures the accumulated definitional divergence from the origin along the path, which aligns conceptually with assessing how well a path maintains coherence relative to its starting point.

Another alternative, summing the SOD scores between consecutive nodes, $\sum_{i=0}^{L-1} \text{SOD}(p_i, p_{i+1})$, was considered less suitable for differentiating between SN and SP paths. Given that navigation steps often involve moving between closely related concepts (especially in SN, but also often in SP due to definition links), the pairwise $\text{SOD}(p_i, p_{i+1})$ is frequently very low (e.g., 1 or 2 if $p_{i+1}$ is directly in the definition of $p_i$). Summing these minimal step-wise scores might result in path scores that are too similar or dominated by path length rather than semantic jumps, potentially obscuring differences between the navigation strategies.

Furthermore, our chosen definition explicitly includes the term $\text{SOD}(p_0, p_L)$, representing the direct definitional distance between the source and the target. While omitting this final term is conceivable, its inclusion ensures that the cumulative score reflects the overall divergence of the entire source-target pair, providing a baseline dimension against which the path's intermediate divergence is accumulated.

Finally, regarding the interpretation of the metric used in the Results Section (Sec.~\ref{sec:results}), a lower value of $\text{SOD}_C(\mathcal{P})$ indicates that the sequence of nodes forming the path $\mathcal{P}$ exhibits less overall definitional divergence from the origin $p_0$. Consequently, paths with lower scores are considered more definitionally coherent or semantically proximal to the source according to the SOD framework.

\subsection{Large-scale experiment design}
\label{subsec:sampling}

To perform a robust comparison between SN and SP using the cumulative SOD metric, and to analyze the relationship between direct SOD and cosine similarity across the different definition processing schemes (as discussed in Sec.~\ref{subsec:sod_vs_cosine_eval}), we conducted a large-scale computational experiment replicated across each of our three dataset conditions (ground-filtered, random removal, targeted removal).

The core experimental design involved generating a large, consistent set of source-target pairs and then computing all relevant metrics for these pairs within the context of each specific dataset condition.

\paragraph*{Source-target pair generation.}
We defined a sampling strategy common to all conditions:
\begin{enumerate}
    \item Fixed set: A curated set of 100 source concepts was selected, aiming to cover a diverse range of semantic domains (e.g., abstract concepts, living beings, objects; see~\cite{ODCodeRepo} for the list). All 100 fixed words were confirmed to be present in our working lexicon.
    \item Complete pool: The pool of potential random concepts consisted of the 19,283 unique entries identified after the initial indexing and ground filtering stages (corresponding to the headwords in the ground-filtered dataset, described in Sec.~\ref{subsec:data_prep}).
    \item Random set selection: A large sample of 10,000 target concepts was randomly selected without replacement from this 19,283-entry pool. This sample represents $51.86\%$ of the eligible lexicon in this range, providing substantial coverage.
    \item Pair generation: Pairs were formed by considering both directions between the 100 fixed words and the 10,000 selected random words, i.e., (fixed, random) and (random, fixed). This resulted in approximately 2 million distinct source-target pairs used for evaluation in each experimental condition.
\end{enumerate}

\paragraph*{Metric computation for each condition.}
Crucially, the subsequent computation of paths and metrics was performed independently for each of the three dataset conditions. For each generated pair $(s, t)$, and for each condition (ground-filtered, random removal, targeted removal), we computed:
\begin{enumerate}
    \item The SN path $\mathcal{P}_{\rm SN}$ and the SP path $\mathcal{Q}_{\rm SP}$, using the specific network graph $G$ and edge weights $w_{ij}$ derived from that condition's definitions (Sec.~\ref{subsec:network}). SN pathfinding utilized the RIM vectors $\mathbf{v}_i$ specific to that condition (Sec.~\ref{subsec:vectors}).
    \item The corresponding cumulative SOD scores, $\text{SOD}_C(\mathcal{P}_{\rm SN})$ and $\text{SOD}_C(\mathcal{Q}_{\rm SP})$, using the SOD framework applied to that condition's definition set (Sec.~\ref{subsec:sod_eval}).
    \item The direct pairwise score $\text{SOD}(s, t)$, again using the definitions from the specific condition.
    \item The direct cosine similarity $\cos(\mathbf{v}_s, \mathbf{v}_t)$, using the RIM vectors $\mathbf{v}_s$ and $\mathbf{v}_t$ specific to that condition.
\end{enumerate}

This comprehensive process resulted in three parallel datasets, each containing the $\sim 2$ million pairs along with their associated navigation paths and SOD/cosine metrics calculated relative to the specific definition processing scheme (ground-filtered, random removal, or targeted removal). These datasets form the basis for the comparative statistical analyses presented in Sec.~\ref{sec:results}.

\subsection{Illustrative example}
\label{subsec:examples}

To further clarify the SOD calculation and its application to path evaluation, we provide an example using the definitions and network structure from the targeted removal dataset condition (Sec.~\ref{subsec:data_prep}).
\subsubsection{Example: path comparison}
\label{ex:path_comparison}

To illustrate how cumulative SOD distinguishes between navigation strategies, consider the paths generated between the source node `zebra' and the target node `circle'. Using SN, which follows maximal cosine similarity to the target vector at each step, the path generated was:
\[ \mathcal{P}_{\rm SN} = (\text{`zebra', `animal', `move', `whirl', `circle'}) \]
Using SP navigation based on Dijkstra's algorithm with edge weights $w_{ij}=1/m_i$, the path found was:
\[ \mathcal{Q}_{\rm SP} = (\text{`zebra', `looks', `full moon', `circle'}) \]

We then calculated the cumulative SOD score for each path using the formula $\text{SOD}_C(\mathcal{P}) = \sum_{i=1}^{L} \text{SOD}(\text{origin}, p_{i})$. The pairwise SOD scores between the origin (`zebra') and each subsequent node on the paths were computed (determining $\omega_{GOD}+1$ for each pair):
\\
\\
\textbf{SN path contributions:}
\begin{center}
\begin{tabular}{lcl}
\text{SOD}(\text{zebra, animal})   & = & 1\\
\text{SOD}(\text{zebra, move})    & =  & 7 \\
\text{SOD}(\text{zebra, whirl})   & = & 40385 \\
\text{SOD}(\text{zebra, circle})  & = & 4659
\end{tabular}
\end{center}
\textbf{SP path contributions:}
\begin{center}
\begin{tabular}{lcl}
    \text{SOD}(\text{zebra, looks})      &=& 1 \\
    \text{SOD}(\text{zebra, full moon})  &=& 388098 \\
    \text{SOD}(\text{zebra, circle})     &=& 4659
\end{tabular}
\end{center}
Summing these contributions yields the total cumulative SOD scores, $\text{SOD}_C(\mathcal{P}_{\rm SN}) = 1 + 7 + 40385 + 4659 = 45052$ and $\text{SOD}_C(\mathcal{Q}_{\rm SP}) = 1 + 388098 + 4659 = 392758$.

In this example, the SN path exhibits a cumulative SOD score nearly an order of magnitude lower than the SP score. This stark difference is primarily driven by the extremely high pairwise $\text{SOD}(\text{zebra, full moon}) = 388098$ encountered along the SP route, signifying a major definitional divergence from the origin (`zebra') introduced by the `full moon' node. While the SP path (`zebra' $\rightarrow$ `looks' $\rightarrow$ `full moon' $\rightarrow$ `circle') is optimal based on structural edge weights (likely exploiting shared terms like `looks'), it appears intuitively less semantically direct than the SN path (`zebra' $\rightarrow$ `animal' $\rightarrow$ `move' $\rightarrow$ `whirl' $\rightarrow$ `circle').

Conversely, the SN path avoids such large definitional jumps relative to the origin, featuring intermediate concepts (`animal', `move', `whirl') that maintain closer definitional relatedness to `zebra'. The cumulative SOD metric quantitatively captures this difference, assigning a lower overall divergence score to the SN path. This example thus illustrates our central hypothesis: the cumulative SOD score, derived purely from definitional structure and independent of the vector space, serves to formalize semantic intuition about path coherence. It provides a quantitative basis for observing that navigation guided by semantic embeddings (SN) tends to produce paths with lower definitional divergence compared to structurally shortest paths (SP), a pattern whose generality we investigate across the large-scale dataset in the following section.

\section{Results}
\label{sec:results}

Using the methodology outlined in Sec.~\ref{sec:application}, involving three distinct definition processing schemes (ground-filtered, random removal, targeted removal), we conducted a large-scale computational analysis on the Simple English Wiktionary network. For each condition, approximately 2 million source-target word pairs were evaluated. The results presented below compare the metrics across these three conditions. We first analyze the relationship between direct SOD scores and RIM cosine similarity, followed by the comparison of cumulative SOD scores for Semantic Navigation (SN) versus Shortest Path (SP) navigation paths. Note that due to path generation failures for some pairs or SOD calculations terminating early based on the $\omega_{GOD}+1$ termination condition for others, the exact number of valid pairs considered may vary slightly between analyses and conditions, but all are approximately $N \approx 2.0 \times 10^6$.

\subsection{Relationship between direct SOD and cosine similarity}
\label{subsec:results_sod_vs_cosine}

Our first analysis investigates the relationship between our proposed definition-based distance metric, the direct pairwise score $\text{SOD}(s, t)$, and a standard measure of semantic similarity derived from network structure, the cosine similarity $\cos(\mathbf{v}_s, \mathbf{v}_t)$ between the RIM vectors. This was performed comparatively across the three dataset conditions.

\begin{figure}[t!] 
    \begin{center}
    \includegraphics[width=0.95\columnwidth]{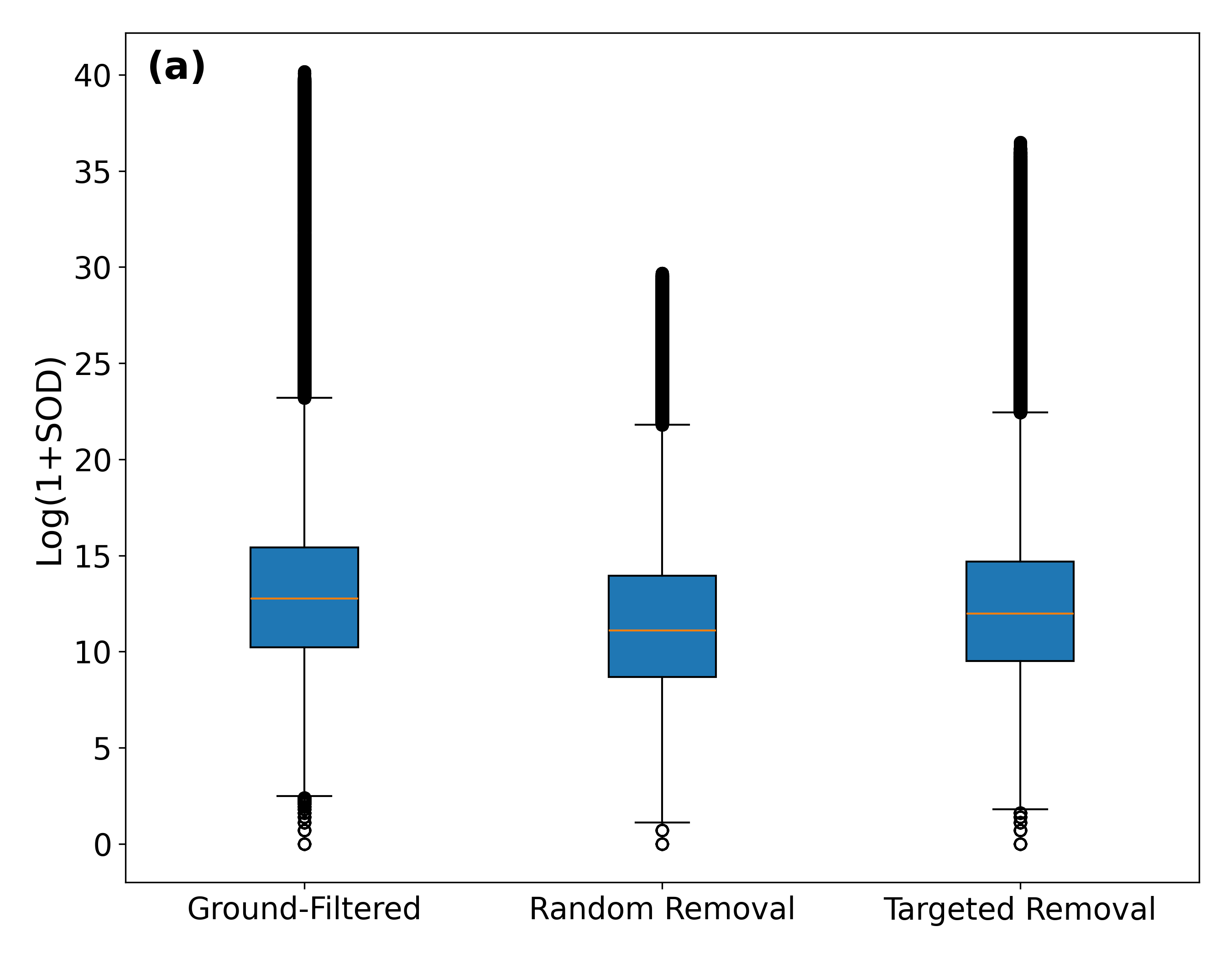} 
    \vspace{1em} 
    \includegraphics[width=0.95\columnwidth]{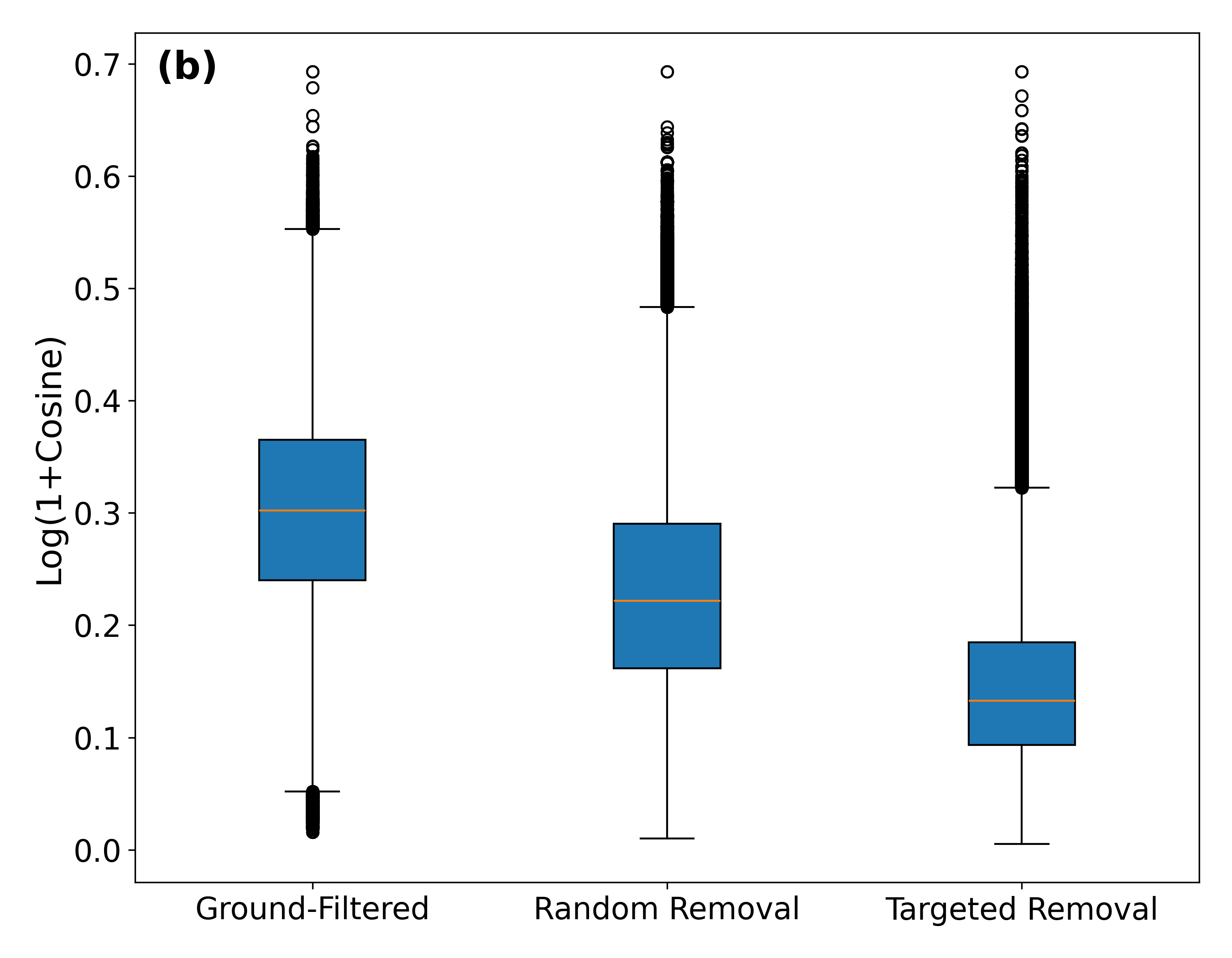} 
    \end{center}
    \caption{Comparative box plots showing the distributions of (a) log(1+SOD) scores and (b) log(1+Cosine) similarity (bottom) across the three dataset conditions: ground-filtered, random removal, and targeted removal. The plots illustrate differences in median (orange line), interquartile range (box), and outlier extent (whiskers and points).}
    \label{fig:boxplots} 
\end{figure}

\paragraph*{Distributions and transformations.}
The raw distributions of the two metrics differ substantially and vary across conditions. Detailed plots showing the distributions at each transformation stage (raw, log(1+x), normalized log(1+x)) for each condition are provided in the Supplementary Material (Figs. S1--S3). Consistently across all conditions, the raw SOD scores exhibit an extremely heavy right tail (see Figs. S1A, S2A, S3A), spanning many orders of magnitude and necessitating transformation for meaningful analysis. Raw cosine similarity values are bounded but also show skewness towards lower values (Figs. S1D, S2D, S3D). Given the extreme skewness, particularly of SOD, we applied a $\log(1+x)$ transformation to both metrics, a standard practice for handling such data. Figure~\ref{fig:boxplots} presents comparative box plots of the resulting log(1+SOD) and log(1+Cosine) distributions across the three conditions. These plots reveal differences in the central tendency and spread; for instance, the median log(1+Cosine) is notably lower for the targeted removal condition compared to the other two, while the log(1+SOD) distributions show variations primarily in their interquartile range and the extent of outliers. Kolmogorov-Smirnov tests confirmed that these log-transformed distributions differ significantly between all pairs of conditions (p $<$ 0.0001; see Sec.~\ref{sec:ks_tests}), underscoring that the definition processing significantly impacts the resulting metric distributions.

\begin{figure*}[t!] 
    \centering
    \includegraphics[width=\textwidth]{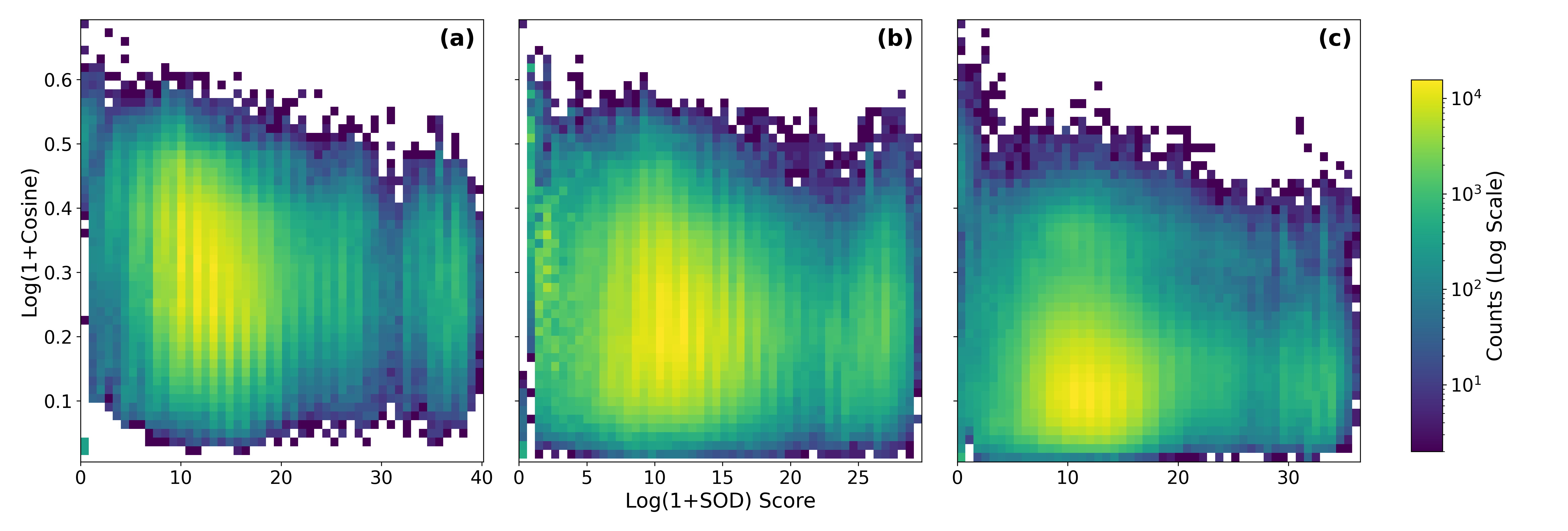} 
    \caption{Comparative 2D density histograms showing the joint distribution of log(1+Cosine) vs. log(1+SOD) for the ground-filtered (a), random removal (b), and targeted removal (c) conditions. Color intensity represents the count of word pairs within each bin on a logarithmic scale. The lack of a clear trend or diagonal structure in all panels visually confirms the weak correlation between the two log-transformed metrics across all conditions.}
    \label{fig:hist2d_comparative}
\end{figure*}

\paragraph*{Visual correlation.}
The relationship between the log-transformed metrics is visualized using comparative 2D density histograms in Fig.~\ref{fig:hist2d_comparative}. Across all three conditions, the plots show dense clouds concentrated at lower log(1+SOD) values, with no discernible linear or non-linear correlation structure. The density appears most concentrated (less diffuse) in the targeted removal condition compared to the ground-filtered and random removal conditions. This visual inspection strongly suggests a weak or negligible correlation between the log-transformed SOD and cosine similarity metrics, regardless of the definition processing scheme.

\paragraph*{Rank correlation.}
To quantify the monotonic relationship between the original metrics, we calculated the Spearman rank correlation coefficient ($\rho$) between the raw $\text{SOD}(s, t)$ scores and the raw $\cos(\mathbf{v}_s, \mathbf{v}_t)$ values for each condition. Table~\ref{tab:spearman_comparative} summarizes these results. A clear trend emerges: the ground-filtered dataset exhibits a weak negative correlation ($\rho \approx -0.21$), which diminishes for the random removal dataset ($\rho \approx -0.10$) and becomes practically negligible for the targeted removal dataset ($\rho \approx -0.03$). While statistically significant due to the large sample sizes ($p < 0.0001$ for all three conditions), even the strongest correlation observed ($-0.21$) is weak in practical terms. This indicates no meaningful rank-order relationship in the targeted removal case, and only a very slight tendency for higher raw SOD scores to correspond with lower raw cosine similarities in the less processed conditions.

\begin{table}[t!] 
    \centering
    \footnotesize 
    \begin{tabular}{l | @{\hspace{0.5em}} c | @{\hspace{0.5em}} c} 
        \hline 
        \textbf{Condition} & \textbf{Valid Pairs} & \textbf{Spearman's $\rho$ (p-value)} \\
        \hline 
        Ground-filtered   & $\approx 2.0 \times 10^6$ & -0.2069 ($p < 0.0001$) \\ 
        Random removal    & $\approx 2.0 \times 10^6$ & -0.0984 ($p < 0.0001$) \\ 
        Targeted removal  & $\approx 2.0 \times 10^6$ & -0.0305 ($p < 0.0001$) \\ 
        \hline 
    \end{tabular}
    \caption{Comparative Spearman rank correlation (raw SOD vs. raw cosine similarity) across conditions.} 
    \label{tab:spearman_comparative}
\end{table}

\begin{figure*}[ht!] 
    \centering
    \includegraphics[width=\textwidth]{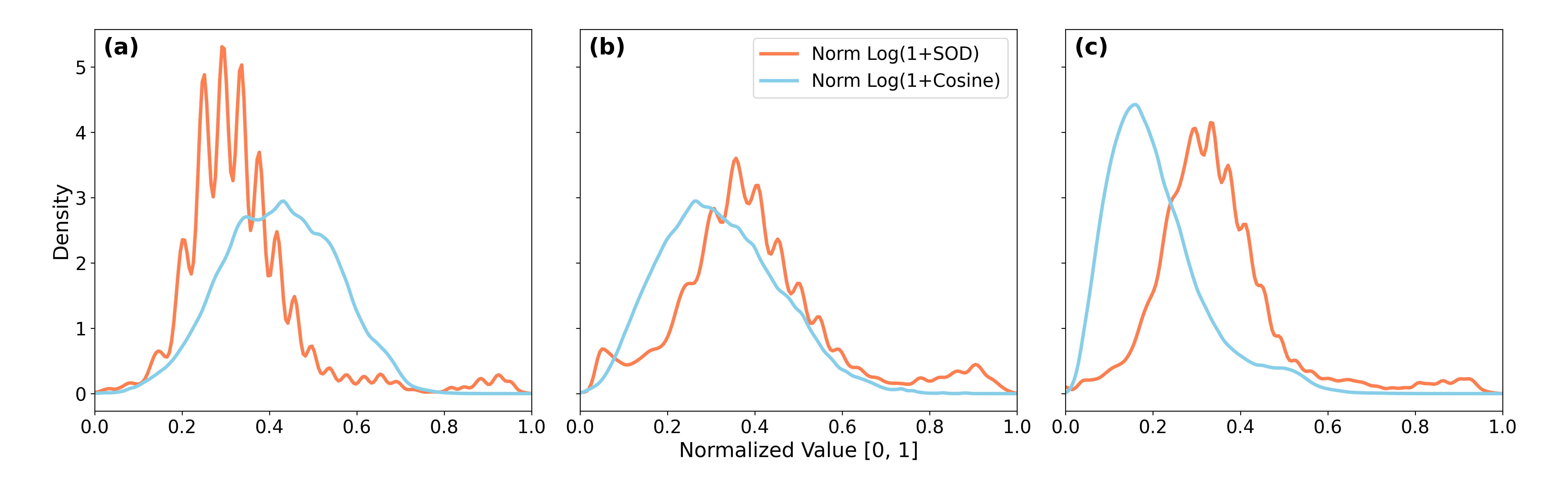}
    \caption{Comparison of the probability density functions for normalized log(1+SOD) and normalized log(1+Cosine) for the ground-filtered (a), random removal (b), and targeted removal (c) conditions. The distinct shapes and peak locations within each panel visually confirm the generally low percentile agreement found between the two metrics' relative rankings (quantified in Table~\ref{tab:percentile_comparative}).}
    \label{fig:norm_log_comparison}
\end{figure*}

\paragraph*{Percentile agreement.}
We further assessed agreement by comparing the relative rankings of pairs after $\log(1+x)$ transformation and min--max normalization, dividing the $[0, 1]$ range into 10 decile bins. Figure~\ref{fig:norm_log_comparison} visually compares the normalized distributions of log(1+SOD) and log(1+Cosine) for each condition, highlighting the general mismatch in shape and peak location between the two metrics within each condition. Table~\ref{tab:percentile_comparative} presents the percentage of pairs falling into the same decile bin for both metrics. Overall agreement is low across all conditions, confirming that the relative ordering of pairs by SOD differs substantially from the ordering by cosine similarity. Interestingly, the agreement is highest for the ground-filtered dataset (17.73\%), followed closely by random removal (17.37\%), and is lowest for the targeted removal dataset (15.07\%).

\begin{table}[t!] 
    \centering
    \footnotesize 
    \begin{tabular}{l | @{\hspace{0.5em}} c | @{\hspace{0.5em}} c} 
        \hline 
        \textbf{Condition} & \textbf{Valid Pairs} & \textbf{Overall Agreement (\%)} \\
        \hline 
        Ground-filtered  & $\approx 2.0 \times 10^6$ & 17.73\% \\ 
        Random removal   & $\approx 2.0 \times 10^6$ & 17.37\% \\ 
        Targeted removal & $\approx 2.0 \times 10^6$ & 15.07\% \\ 
        \hline 
    \end{tabular}
    \caption{Comparative percentile agreement analysis: normalized log(1+SOD) vs. normalized log(1+Cosine) (10 Bins).} 
    \label{tab:percentile_comparative}
\end{table}

\paragraph*{Conceptual comparison through illustrative examples.}
To provide a more intuitive and qualitative understanding of this orthogonality, we performed a focused analysis on a curated set of nine concepts: `dog', `moon', `sea', `doctor', `war', `hospital', `police', `soldier', and `circus'. This set was specifically chosen because it contains pairs with a priori intuitive semantic relationships of varying degrees. This allows us to evaluate how well each metric's similarity ranking aligns with human intuition. Before presenting the results, we can hypothesize a more nuanced hierarchy of relationships:
\begin{itemize}
    \item \textbf{First-order relationships:} Strong, direct connections such as (`doctor', `hospital'), (`war', `soldier'), and the physical/conceptual link (`moon', `sea'). A robust semantic metric should rank these highest.
    \item \textbf{Second-order relationships:} Clear but less direct links, including functional associations like (`soldier', `police'), (`dog', `police'), (`sea', `police'), or thematic connections like (`doctor', `war').
    \item \textbf{Third-order relationships:} More abstract or indirect links, such as (`war', `hospital') and (`doctor', `police').
\end{itemize}

We computed the direct SOD score and cosine similarity for all 36 unique combinations of these words using the ground-filtered dataset. The top five and bottom five most similar pairs, as ranked by each metric, are presented in Table~\ref{tab:sod_cosine_examples}.

\begin{table}[h!]
    \centering
    
    \setlength{\tabcolsep}{1.5em} 
    
    \begin{tabular}{cll}
        \toprule
        \textbf{Rank} & \multicolumn{1}{c}{\textbf{SOD}} & \multicolumn{1}{c}{\textbf{Cosine}} \\
        \midrule
        1 & (doctor, hospital) & (doctor, hospital) \\
        2 & (war, soldier)     & (moon, sea) \\
        3 & (moon, sea)        & (war, police) \\
        4 & (sea, police)      & (doctor, police) \\
        5 & (doctor, war)      & (hospital, police) \\

        & \multicolumn{2}{c}{\vdots} \\
        
        32 & (dog, soldier) & (war, circus) \\
        33 & (dog, war)   & (soldier, circus) \\
        34 & (dog, circus)       & (dog, circus) \\
        35 & (soldier, circus)    & (dog, soldier) \\
        36 & (war, circus) & (dog, hospital) \\
        \bottomrule
    \end{tabular}
    \caption{
        The five most similar (top) and five least similar (bottom) word pairs from a curated set of 36, as ranked by SOD and Cosine Similarity.
    }
    \label{tab:sod_cosine_examples}    
\end{table}

The rankings in Table~\ref{tab:sod_cosine_examples} reveal the distinct nature of the metrics and their alignment with our intuitive hierarchy. The SOD rankings are remarkably consistent with our a priori expectations. Its top three pairs---(`doctor', `hospital'), (`war', `soldier'), and (`moon', `sea')---are precisely the three we identified as having first-order relationships. Furthermore, its top five is rounded out by the strong second-order pairs (`sea', `police') and (`doctor', `war'). This demonstrates SOD's effectiveness in capturing structurally significant definitional links. In contrast, while cosine similarity also identifies (`doctor', `hospital') and (`moon', `sea') in its top two, it fails to recognize the salient `war-soldier' connection, which does not appear in its top five. Instead, its rankings are heavily influenced by thematic proximity, with three of its top five pairs involving `police' in combination with other civil or emergency-service concepts.

For pairs not captured in the top five, the results are more mixed and further highlight the trade-offs between the metrics. The third-order pair (`war', `hospital') is ranked similarly by both but still better by SOD (14th by SOD, 16th by cosine). However, cosine similarity aligns better with some of our hypothesized second-order functional relationships, ranking (`soldier', `police') 7th, whereas SOD ranks it a distant 25th. Similarly, (`dog', `police') is ranked 24th by cosine versus 29th by SOD. This reinforces the idea that cosine is highly adept at capturing thematic and functional associations derived from contextual usage.

Looking at the bottom of the rankings, we observe a strong consensus. Both metrics agree that pairs involving `circus' are semantically distant from the other concepts, and both rank the pair `dog-soldier' as highly dissimilar. This concordance on clearly unrelated pairs validates that both metrics are behaving sensibly. The superior alignment of SOD's top rankings with our intuitively structured hypothesis, however, highlights its strength as a measure of definition-grounded semantics. These illustrative examples substantiate our large-scale findings, confirming that SOD measures semantic proximity through definitional overlap, while cosine similarity measures it through contextual association.

\paragraph*{Summary of SOD vs. cosine comparison.}
Taken together, our large-scale quantitative and focused qualitative analyses consistently demonstrate a weak relationship between direct SOD scores and RIM cosine similarity. The low rank correlation and percentile agreement across millions of pairs indicate statistical orthogonality, and our illustrative examples provide the conceptual grounding for this finding. The qualitative comparison revealed that SOD's rankings align remarkably well with a priori definitional hierarchies, successfully identifying core relationships such as (`war', `soldier') that are missed by cosine similarity. Conversely, cosine excels at identifying the thematic and contextual associations for pairs like (`soldier', `police') and (`dog', `police'). This trend, which becomes more pronounced as definitions are curated, confirms that SOD captures a distinct aspect of semantic relatedness grounded in explicit definitional structure, which is largely independent of the network proximity and contextual information captured by RIM cosine similarity.

\subsection{Comparison of cumulative SOD for SN vs. SP paths}
\label{subsec:results_path_comparison}

Next, we evaluated the semantic coherence of paths generated by the SN and SP algorithms by comparing their cumulative SOD scores. For a given path $\mathcal{P}$, this score is defined as $\text{SOD}_C(\mathcal{P}) = \sum_{i=1}^{L} \text{SOD}(p_0, p_{i})_{\omega_{GOD}+1}$ (see Sec.~\ref{subsec:sod_eval}). This evaluation was performed across the ~2 million source-target pairs for each of the three dataset conditions. A lower $\text{SOD}_C$ score indicates greater definitional coherence relative to the path origin.

\begin{table}[t!] 
    \centering
    \footnotesize 
    \newcommand{\mycolwidth}{0.22\columnwidth} 

    \begin{tabular}{l | >{\centering\arraybackslash}p{\mycolwidth} | >{\centering\arraybackslash}p{\mycolwidth} | >{\centering\arraybackslash}p{\mycolwidth}} 
        \hline 
        \textbf{Condition} & \textbf{SN $<$ SP} & \textbf{SN $=$ SP} & \textbf{SN $>$ SP} \\ 
        \hline 
        Ground-filtered  & 59.82\% & 16.96\% & 23.22\% \\ 
        Random removal   & 62.42\% & 14.80\% & 22.78\% \\ 
        Targeted removal & 58.22\% & 15.48\% & 26.30\% \\ 
        \hline 
    \end{tabular}
    \caption{Comparison of path outcomes based on cumulative SOD scores ($\text{SOD}_C$) for semantic navigation (SN) vs. shortest path (SP) across conditions. Headers indicate the comparison of $\text{SOD}_C(\text{SN})$ relative to $\text{SOD}_C(\text{SP})$.}
    \label{tab:sn_sp_comparison}
\end{table}

\paragraph*{Dominance of semantic navigation.}
Table~\ref{tab:sn_sp_comparison} presents the aggregated results comparing the $\text{SOD}_C$ scores for paths generated by SN versus those by SP between the same endpoints for each condition. A consistent and robust finding emerges across all three processing schemes: paths generated by SN exhibit lower $\text{SOD}_C$ scores than paths generated by SP in a significant majority of cases (59.82\% for ground-filtered, 62.42\% for random removal, and 58.22\% for targeted removal). This primary observation holds regardless of the definition processing method applied. Statistical tests further confirm the significance of this dominance within each condition: $\chi^2$ goodness-of-fit tests strongly reject the hypothesis of equal likelihood for the three outcomes, namely: $\text{SOD}_C(\text{SN}) < \text{SOD}_C(\text{SP})$, $\text{SOD}_C(\text{SN}) = \text{SOD}_C(\text{SP})$, $\text{SOD}_C(\text{SN}) > \text{SOD}_C(\text{SP})$ ($p<$ 0.0001 for all conditions). Furthermore, one-sided binomial tests confirm that the proportion of pairs where $\text{SOD}_C(\text{SN}) < \text{SOD}_C(\text{SP})$ is significantly greater than 50\% for each condition ($p<$ 0.0001). This provides strong evidence that SN consistently generates paths more definitionally coherent with the origin (as measured by lower $\text{SOD}_C$) compared to SP, demonstrating the robustness of using this cumulative SOD metric for validating semantic navigation across different corpus characteristics. This fundamental result is further tested and confirmed on more restrictive directed networks in the SM~\cite{SM},
where we also control for the confounding effect of path length.

\paragraph*{Sensitivity of comparison outcomes to network conditions.}
While the core finding—that SN paths are generally favored by the $\text{SOD}_C$ metric—is robust, the precise distribution of the three comparison outcomes, $\text{SOD}_C(\text{SN})$ relative to $\text{SOD}_C(\text{SP})$, does show statistically significant variation across the different network conditions. This is confirmed by a $\chi^2$ test for contingency applied to the counts underlying Table~\ref{tab:sn_sp_comparison} ($\chi^2 = 12156.70$, $\text{df}=4$, $p < 0.0001$), indicating that the definition processing scheme influences the quantitative results of the SN vs. SP comparison. Notably, the random removal condition shows the highest proportion of cases where $\text{SOD}_C(\text{SN}) < \text{SOD}_C(\text{SP})$ (62.42\%). Conversely, the targeted removal condition exhibits the lowest proportion in this category (58.22\%) and the highest proportion where $\text{SOD}_C(\text{SN}) > \text{SOD}_C(\text{SP})$ (26.30\%). This suggests that the specific structural changes induced by different processing methods (e.g., random noise vs. targeted removal of common words) can subtly alter the relative performance of SN and SP algorithms when evaluated using the definition-based $\text{SOD}_C$ score. For instance, removing random links might disproportionately affect the structurally optimal SP paths, enhancing SN's relative advantage, whereas targeted removal might occasionally create shortcuts usable by SP that are also definitionally sound. These variations highlight the interplay between network structure and navigation algorithm performance, but they do not undermine the consistent overall observation that SN paths align better with definitional coherence as measured by cumulative SOD.

\subsection{Statistical comparison of metric distributions}
\label{sec:ks_tests}

To further quantify the impact of the definition processing schemes, we performed two-sample Kolmogorov-Smirnov (KS) tests comparing the distributions of the log-transformed metrics between pairs of conditions. The results (Table~\ref{tab:ks_tests}) show statistically significant differences ($p<$ 0.0001) for both log(1+SOD) and log(1+Cosine) distributions between all pairs of conditions (ground-filtered vs. random removal, ground-filtered vs. targeted removal, and random removal vs. targeted removal). These statistical differences align with the visual distinctions observed in the comparative box plots (Fig.~\ref{fig:boxplots}). This confirms that the different definition processing strategies indeed result in measurably distinct statistical properties for both the SOD and cosine similarity metrics derived from the resulting networks.

\begin{table}[t!] 
    \centering
    \footnotesize 
    \newcommand{\ksmetricwidthSmall}{0.25\columnwidth} 
    \newcommand{\ksstatwidthSmall}{0.35\columnwidth}  
    \begin{tabular}{l | >{\centering\arraybackslash}p{\ksmetricwidthSmall} | >{\centering\arraybackslash}p{\ksstatwidthSmall}} 
        \hline 
        \textbf{Comparison} & \textbf{Metric} & \textbf{KS Stat (p-value)} \\ 
        \hline 
        GF vs RR  & log(1+SOD)   & 0.1842 ($p \ll 0.0001$) \\ 
                  & log(1+Cosine) & 0.3216 ($p \ll 0.0001$) \\ \hline
        GF vs TR  & log(1+SOD)   & 0.1006 ($p \ll 0.0001$) \\ 
                  & log(1+Cosine) & 0.6760 ($p \ll 0.0001$) \\ \hline
        RR vs TR  & log(1+SOD)   & 0.0864 ($p \ll 0.0001$) \\ 
                  & log(1+Cosine) & 0.4088 ($p \ll 0.0001$) \\
        \hline 
    \end{tabular}
    \caption{Kolmogorov-Smirnov (KS) 2-sample test results comparing log(1+SOD) and log(1+Cosine) distributions between conditions: ground-filtered (GF), random removal (RR), targeted removal (TR).} 
    \label{tab:ks_tests}
\end{table}

\section{Conclusions}
\label{sec:conclusions}

Quantifying semantic relationships directly from the structure of lexical networks remains a fundamental challenge. While vector embedding techniques capture contextual proximity, intrinsic measures reflecting explicit definitional relationships are less explored. In this work, we introduced and formalized Ontological Differentiation (OD), specifically Strong Ontological Differentiation (SOD), as a novel method to compute a divergence score based purely on the recursive expansion and overlap of their definitions. We aimed to demonstrate its utility as a measure of semantic structure and coherence within these networks. To establish the robustness and general applicability of our findings beyond the specifics of any single corpus representation, we performed our analyses comparatively across three variations of a network derived from the Simple English Wiktionary, created using distinct definition processing schemes (ground-filtered, random removal, and targeted removal) to simulate diverse corpus characteristics.

Our investigation yielded several key conclusions. First, we examined the relationship between direct pairwise SOD scores and RIM cosine similarity. The results revealed that this relationship is sensitive to definition processing: a weak negative correlation in the ground-filtered network ($\rho \approx -0.21$) progressively diminished to become negligible in the targeted removal network ($\rho \approx -0.03$) (Table~\ref{tab:spearman_comparative}). This trend, supported by percentile agreement analyses (Table~\ref{tab:percentile_comparative}), establishes that SOD is largely orthogonal to embedding-based similarity. This statistical independence is grounded in a fundamental conceptual distinction, as highlighted by our illustrative examples (Table~\ref{tab:sod_cosine_examples}). The analysis revealed that SOD excels at capturing explicit definitional hierarchies, successfully identifying core relationships such as (`war', `soldier') that are missed by cosine similarity's top rankings. Conversely, cosine similarity proved more adept at identifying thematic and contextual associations, such as (`soldier', `police'). This confirms that SOD offers a complementary, definition-grounded perspective on semantic relatedness that is distinct from existing vector-space and gloss-overlap methods.

Second, we utilized the cumulative SOD score ($\text{SOD}_C$) as an independent benchmark to evaluate the semantic coherence of paths generated by semantic navigation versus shortest path algorithms. This core evaluation, replicated across all three network conditions, demonstrated the utility and robustness of the $\text{SOD}_C$ metric. We found that SN paths consistently exhibited significantly lower $\text{SOD}_C$ scores compared to SP paths across all three datasets (Table~\ref{tab:sn_sp_comparison}), indicating greater definitional coherence relative to the origin. This finding was proved to be statistically robust across all three network conditions ---whether the definitions were minimally processed, randomly perturbed, or specifically curated--- with one-sided binomial tests for each confirming that SN paths outperform SP paths significantly more often than the 50\% chance level ($p < 0.0001$). The consistency of this result across the simulated variations in corpus structure provides strong evidence that cumulative SOD serves as a generally effective metric for validating the semantic coherence of navigation processes relative to definitional structure. While the precise magnitude of SN's advantage varied significantly between conditions (as shown by a $\chi^2$ test, $p<$ 0.0001), its overall superiority as measured by $\text{SOD}_C$ remained constant.

This finding—that SN paths consistently exhibit lower cumulative SOD scores than SP paths—offers a key insight into the nature of navigation in semantic networks. This result highlights a fundamental distinction between two navigation philosophies: structural economy versus semantic coherence. The Shortest Path (SP) algorithm, by design, optimizes for structural economy, identifying the most direct path available through the explicit definitional links. In contrast, Semantic Navigation (SN) uses vector embeddings that encode high-order, network-wide relationships to maintain coherence with the target. The fact that SOD, a metric grounded purely in the dictionary's explicit definitional structure, consistently favors SN paths is a non-trivial result. It provides strong quantitative evidence that the statistical patterns learned by the RIM embedding method successfully capture the underlying semantic coherence of the lexicon. In other words, SN's greedy, vector-guided strategy produces trajectories that are not only statistically plausible but are also more faithful to the formal meaning encoded in dictionary definitions. This validates the use of such embeddings for generating semantically meaningful pathways and demonstrates that SOD can serve as an effective, independent benchmark for measuring this alignment between statistical and symbolic representations of knowledge.

Beyond its role as an evaluative metric, SOD can be regarded as a constructive heuristic for semantic navigation itself. In the Section II of the SM~\cite{SM}, we introduce a proof-of-concept for semantic navigation based on SOD (which we term Ontological Navigation, ON), based on greedily selecting paths by minimizing the direct SOD score to the target. Qualitative analysis of the resulting paths (see Supplemental Table~S1) shows they are often qualitatively different from those of SN and follow a more intuitive, definition-based logic, exemplified by the path `dog' $\rightarrow$ `feist' $\rightarrow$ `belligerent' $\rightarrow$ `war'. This demonstrates that SOD is not merely a passive validation tool but can serve as an active and principled heuristic, opening a new avenue for purely symbolic semantic navigation.

In essence, Ontological Differentiation provides a formal method grounded in shared definitional components. Our comparative analysis across systematically varied definition sets validates SOD as a definition-based measure of semantic divergence, distinct from vector similarity. Most importantly, its consistent ability to identify SN paths as more definitionally coherent than SP paths across varied network structures demonstrates its robustness and utility as a tool for analyzing and validating semantic processes in lexical networks, offering insights independent of specific corpus construction details. Furthermore, this work demonstrates its dual potential: beyond its power as an analytical tool, its promise as a constructive heuristic is introduced through our proof-of-concept Ontological Navigation. Our analysis showing that SOD is coherently sensitive to polysemy suggests promising future extensions for this navigation, potentially allowing for context-aware pathfinding. The conclusions drawn from our main analysis on symmetric networks are decisively reinforced by our extended analysis in the SM~\cite{SM}.
There, we show that the superior coherence of SN paths holds even on structurally constrained directed networks once the confounding effect of path length is properly controlled, thereby validating SOD's capacity to serve as a robust evaluation tool whose findings are independent of specific corpus construction or network topology. The principled nature of the framework, supported by this demonstrated robustness, also points to its generalizability beyond the linguistic domain to mathematical and physical ontologies, a subject of ongoing investigation.

\section*{Acknowledgments}

Funding was provided by the PRIORITY grant (PID2021-127202NB-C22) to JAC, and has been partially supported by the grant PID2021-122711NB-C21, both funded by MCIN / AEI / 10.13039 / 501100011033 and ``ERDF A way of making Europe''. 

The authors acknowledge computing resources at the Magerit Supercomputer of the
Universidad Polit\'{e}cnica de Madrid.

\bigskip


\bibliography{myreferences}

\begin{thebibliography}{24}%
\makeatletter
\providecommand \@ifxundefined [1]{%
 \@ifx{#1\undefined}
}%
\providecommand \@ifnum [1]{%
 \ifnum #1\expandafter \@firstoftwo
 \else \expandafter \@secondoftwo
 \fi
}%
\providecommand \@ifx [1]{%
 \ifx #1\expandafter \@firstoftwo
 \else \expandafter \@secondoftwo
 \fi
}%
\providecommand \natexlab [1]{#1}%
\providecommand \enquote  [1]{``#1''}%
\providecommand \bibnamefont  [1]{#1}%
\providecommand \bibfnamefont [1]{#1}%
\providecommand \citenamefont [1]{#1}%
\providecommand \href@noop [0]{\@secondoftwo}%
\providecommand \href [0]{\begingroup \@sanitize@url \@href}%
\providecommand \@href[1]{\@@startlink{#1}\@@href}%
\providecommand \@@href[1]{\endgroup#1\@@endlink}%
\providecommand \@sanitize@url [0]{\catcode `\\12\catcode `\$12\catcode `\&12\catcode `\#12\catcode `\^12\catcode `\_12\catcode `\%12\relax}%
\providecommand \@@startlink[1]{}%
\providecommand \@@endlink[0]{}%
\providecommand \url  [0]{\begingroup\@sanitize@url \@url }%
\providecommand \@url [1]{\endgroup\@href {#1}{\urlprefix }}%
\providecommand \urlprefix  [0]{URL }%
\providecommand \Eprint [0]{\href }%
\providecommand \doibase [0]{https://doi.org/}%
\providecommand \selectlanguage [0]{\@gobble}%
\providecommand \bibinfo  [0]{\@secondoftwo}%
\providecommand \bibfield  [0]{\@secondoftwo}%
\providecommand \translation [1]{[#1]}%
\providecommand \BibitemOpen [0]{}%
\providecommand \bibitemStop [0]{}%
\providecommand \bibitemNoStop [0]{.\EOS\space}%
\providecommand \EOS [0]{\spacefactor3000\relax}%
\providecommand \BibitemShut  [1]{\csname bibitem#1\endcsname}%
\let\auto@bib@innerbib\@empty
\bibitem [{\citenamefont {Ferrer~i Cancho}\ and\ \citenamefont {Sol{\'e}}(2001)}]{CanchoFerrerSole2001}%
  \BibitemOpen
  \bibfield  {author} {\bibinfo {author} {\bibfnamefont {R.}~\bibnamefont {Ferrer~i Cancho}}\ and\ \bibinfo {author} {\bibfnamefont {R.~V.}\ \bibnamefont {Sol{\'e}}},\ }\href {https://doi.org/10.1098/rspb.2001.1800} {\bibfield  {journal} {\bibinfo  {journal} {Proceedings of the Royal Society of London. Series B: Biological Sciences}\ }\textbf {\bibinfo {volume} {268}},\ \bibinfo {pages} {1482} (\bibinfo {year} {2001})}\BibitemShut {NoStop}%
\bibitem [{\citenamefont {Resnik}(1995)}]{Resnik1995}%
  \BibitemOpen
  \bibfield  {author} {\bibinfo {author} {\bibfnamefont {P.}~\bibnamefont {Resnik}},\ }in\ \href@noop {} {\emph {\bibinfo {booktitle} {Proceedings of the 14th International Joint Conference on Artificial Intelligence (IJCAI'95)}}},\ Vol.~\bibinfo {volume} {1}\ (\bibinfo {organization} {Morgan Kaufmann Publishers Inc.},\ \bibinfo {year} {1995})\ pp.\ \bibinfo {pages} {448--453}\BibitemShut {NoStop}%
\bibitem [{\citenamefont {Boccaletti}\ \emph {et~al.}(2006)\citenamefont {Boccaletti}, \citenamefont {Latora}, \citenamefont {Moreno}, \citenamefont {Chavez},\ and\ \citenamefont {Hwang}}]{BoccalettiEtAl2006}%
  \BibitemOpen
  \bibfield  {author} {\bibinfo {author} {\bibfnamefont {S.}~\bibnamefont {Boccaletti}}, \bibinfo {author} {\bibfnamefont {V.}~\bibnamefont {Latora}}, \bibinfo {author} {\bibfnamefont {Y.}~\bibnamefont {Moreno}}, \bibinfo {author} {\bibfnamefont {M.}~\bibnamefont {Chavez}},\ and\ \bibinfo {author} {\bibfnamefont {D.~U.}\ \bibnamefont {Hwang}},\ }\href {https://doi.org/10.1016/j.physrep.2005.10.009} {\bibfield  {journal} {\bibinfo  {journal} {Physics Reports}\ }\textbf {\bibinfo {volume} {424}},\ \bibinfo {pages} {175} (\bibinfo {year} {2006})}\BibitemShut {NoStop}%
\bibitem [{\citenamefont {Sigman}\ and\ \citenamefont {Cecchi}(2002)}]{SigmanCecchi2002}%
  \BibitemOpen
  \bibfield  {author} {\bibinfo {author} {\bibfnamefont {M.}~\bibnamefont {Sigman}}\ and\ \bibinfo {author} {\bibfnamefont {G.~A.}\ \bibnamefont {Cecchi}},\ }\href {https://doi.org/10.1073/pnas.022341799} {\bibfield  {journal} {\bibinfo  {journal} {Proceedings of the National Academy of Sciences}\ }\textbf {\bibinfo {volume} {99}},\ \bibinfo {pages} {1742} (\bibinfo {year} {2002})}\BibitemShut {NoStop}%
\bibitem [{\citenamefont {Motter}\ \emph {et~al.}(2002)\citenamefont {Motter}, \citenamefont {De~Moura}, \citenamefont {Lai},\ and\ \citenamefont {Dasgupta}}]{MotterEtAl2002}%
  \BibitemOpen
  \bibfield  {author} {\bibinfo {author} {\bibfnamefont {A.~E.}\ \bibnamefont {Motter}}, \bibinfo {author} {\bibfnamefont {A.~P.~S.}\ \bibnamefont {De~Moura}}, \bibinfo {author} {\bibfnamefont {Y.-C.}\ \bibnamefont {Lai}},\ and\ \bibinfo {author} {\bibfnamefont {P.}~\bibnamefont {Dasgupta}},\ }\href {https://doi.org/10.1103/PhysRevE.65.065102} {\bibfield  {journal} {\bibinfo  {journal} {Physical Review E}\ }\textbf {\bibinfo {volume} {65}},\ \bibinfo {pages} {065102} (\bibinfo {year} {2002})}\BibitemShut {NoStop}%
\bibitem [{\citenamefont {Steyvers}\ and\ \citenamefont {Tenenbaum}(2005)}]{SteyversTenenbaum2005}%
  \BibitemOpen
  \bibfield  {author} {\bibinfo {author} {\bibfnamefont {M.}~\bibnamefont {Steyvers}}\ and\ \bibinfo {author} {\bibfnamefont {J.~B.}\ \bibnamefont {Tenenbaum}},\ }\href {https://doi.org/10.1207/s15516709cog2901_3} {\bibfield  {journal} {\bibinfo  {journal} {Cognitive Science}\ }\textbf {\bibinfo {volume} {29}},\ \bibinfo {pages} {41} (\bibinfo {year} {2005})}\BibitemShut {NoStop}%
\bibitem [{\citenamefont {Vitevitch}(2008)}]{Vitevitch2008}%
  \BibitemOpen
  \bibfield  {author} {\bibinfo {author} {\bibfnamefont {M.~S.}\ \bibnamefont {Vitevitch}},\ }\href@noop {} {\bibfield  {journal} {\bibinfo  {journal} {Journal of Speech, Language, and Hearing Research}\ }\textbf {\bibinfo {volume} {51}},\ \bibinfo {pages} {408} (\bibinfo {year} {2008})}\BibitemShut {NoStop}%
\bibitem [{\citenamefont {Bordes}\ \emph {et~al.}(2014)\citenamefont {Bordes}, \citenamefont {Weston}, \citenamefont {Chopra},\ and\ \citenamefont {Collobert}}]{BordesWeston2014}%
  \BibitemOpen
  \bibfield  {author} {\bibinfo {author} {\bibfnamefont {A.}~\bibnamefont {Bordes}}, \bibinfo {author} {\bibfnamefont {J.}~\bibnamefont {Weston}}, \bibinfo {author} {\bibfnamefont {S.}~\bibnamefont {Chopra}},\ and\ \bibinfo {author} {\bibfnamefont {R.}~\bibnamefont {Collobert}},\ }in\ \href {https://doi.org/10.3115/v1/D14-1067} {\emph {\bibinfo {booktitle} {Proceedings of the 2014 Conference on Empirical Methods in Natural Language Processing (EMNLP)}}}\ (\bibinfo  {publisher} {Association for Computational Linguistics},\ \bibinfo {address} {Doha, Qatar},\ \bibinfo {year} {2014})\ pp.\ \bibinfo {pages} {615--620}\BibitemShut {NoStop}%
\bibitem [{\citenamefont {Nickel}\ and\ \citenamefont {Kiela}(2017)}]{NickelKiela2017}%
  \BibitemOpen
  \bibfield  {author} {\bibinfo {author} {\bibfnamefont {M.}~\bibnamefont {Nickel}}\ and\ \bibinfo {author} {\bibfnamefont {D.}~\bibnamefont {Kiela}},\ }in\ \href {http://papers.nips.cc/paper/7213-poincare-embeddings-for-learning-hierarchical-representations.pdf} {\emph {\bibinfo {booktitle} {Advances in Neural Information Processing Systems 30 (NIPS 2017)}}},\ \bibinfo {series and number} {\bibinfo {number} {30}},\ \bibinfo {editor} {edited by\ \bibinfo {editor} {\bibfnamefont {I.}~\bibnamefont {Guyon}}, \bibinfo {editor} {\bibfnamefont {U.~V.}\ \bibnamefont {Luxburg}}, \bibinfo {editor} {\bibfnamefont {S.}~\bibnamefont {Bengio}}, \bibinfo {editor} {\bibfnamefont {H.}~\bibnamefont {Wallach}}, \bibinfo {editor} {\bibfnamefont {R.}~\bibnamefont {Fergus}}, \bibinfo {editor} {\bibfnamefont {S.}~\bibnamefont {Vishwanathan}},\ and\ \bibinfo {editor} {\bibfnamefont {R.}~\bibnamefont {Garnett}}}\ (\bibinfo  {publisher} {Curran Associates, Inc.},\ \bibinfo {year} {2017})\BibitemShut {NoStop}%
\bibitem [{\citenamefont {Perozzi}\ \emph {et~al.}(2014)\citenamefont {Perozzi}, \citenamefont {Al-Rfou},\ and\ \citenamefont {Skiena}}]{PerozziAlRfou2014}%
  \BibitemOpen
  \bibfield  {author} {\bibinfo {author} {\bibfnamefont {B.}~\bibnamefont {Perozzi}}, \bibinfo {author} {\bibfnamefont {R.}~\bibnamefont {Al-Rfou}},\ and\ \bibinfo {author} {\bibfnamefont {S.}~\bibnamefont {Skiena}},\ }in\ \href {https://doi.org/10.1145/2623330.2623732} {\emph {\bibinfo {booktitle} {Proceedings of the 20th ACM SIGKDD International Conference on Knowledge Discovery and Data Mining (KDD '14)}}}\ (\bibinfo  {publisher} {Association for Computing Machinery},\ \bibinfo {address} {New York, NY, USA},\ \bibinfo {year} {2014})\ pp.\ \bibinfo {pages} {701--710}\BibitemShut {NoStop}%
\bibitem [{\citenamefont {Mikolov}\ \emph {et~al.}(2013)\citenamefont {Mikolov}, \citenamefont {Sutskever}, \citenamefont {Chen}, \citenamefont {Corrado},\ and\ \citenamefont {Dean}}]{MikolovEtAl2013}%
  \BibitemOpen
  \bibfield  {author} {\bibinfo {author} {\bibfnamefont {T.}~\bibnamefont {Mikolov}}, \bibinfo {author} {\bibfnamefont {I.}~\bibnamefont {Sutskever}}, \bibinfo {author} {\bibfnamefont {K.}~\bibnamefont {Chen}}, \bibinfo {author} {\bibfnamefont {G.~S.}\ \bibnamefont {Corrado}},\ and\ \bibinfo {author} {\bibfnamefont {J.}~\bibnamefont {Dean}},\ }in\ \href {http://papers.nips.cc/paper/5021-distributed-representations-of-words-and-phrases-and-their-compositionality.pdf} {\emph {\bibinfo {booktitle} {Advances in Neural Information Processing Systems 26 (NIPS 2013)}}},\ \bibinfo {editor} {edited by\ \bibinfo {editor} {\bibfnamefont {C.~J.~C.}\ \bibnamefont {Burges}}, \bibinfo {editor} {\bibfnamefont {L.}~\bibnamefont {Bottou}}, \bibinfo {editor} {\bibfnamefont {M.}~\bibnamefont {Welling}}, \bibinfo {editor} {\bibfnamefont {Z.}~\bibnamefont {Ghahramani}},\ and\ \bibinfo {editor} {\bibfnamefont {K.~Q.}\ \bibnamefont {Weinberger}}}\ (\bibinfo  {publisher} {Curran Associates, Inc.},\ \bibinfo {year} {2013})\BibitemShut
  {NoStop}%
\bibitem [{\citenamefont {Pennington}\ \emph {et~al.}(2014)\citenamefont {Pennington}, \citenamefont {Socher},\ and\ \citenamefont {Manning}}]{PenningtonEtAl2014}%
  \BibitemOpen
  \bibfield  {author} {\bibinfo {author} {\bibfnamefont {J.}~\bibnamefont {Pennington}}, \bibinfo {author} {\bibfnamefont {R.}~\bibnamefont {Socher}},\ and\ \bibinfo {author} {\bibfnamefont {C.~D.}\ \bibnamefont {Manning}},\ }in\ \href {https://doi.org/10.3115/v1/D14-1162} {\emph {\bibinfo {booktitle} {Proceedings of the 2014 Conference on Empirical Methods in Natural Language Processing (EMNLP)}}}\ (\bibinfo  {publisher} {Association for Computational Linguistics},\ \bibinfo {address} {Doha, Qatar},\ \bibinfo {year} {2014})\ pp.\ \bibinfo {pages} {1532--1543}\BibitemShut {NoStop}%
\bibitem [{\citenamefont {Borge-Holthoefer}\ and\ \citenamefont {Arenas}(2010)}]{BorgeHolthoeferArenas2010}%
  \BibitemOpen
  \bibfield  {author} {\bibinfo {author} {\bibfnamefont {J.}~\bibnamefont {Borge-Holthoefer}}\ and\ \bibinfo {author} {\bibfnamefont {A.}~\bibnamefont {Arenas}},\ }\href {https://epjb.epj.org/articles/epjb/abs/2010/06/b090804/b090804.html} {\bibfield  {journal} {\bibinfo  {journal} {The European Physical Journal B}\ }\textbf {\bibinfo {volume} {74}},\ \bibinfo {pages} {265} (\bibinfo {year} {2010})}\BibitemShut {NoStop}%
\bibitem [{\citenamefont {Fouss}\ \emph {et~al.}(2007)\citenamefont {Fouss}, \citenamefont {Pirotte}, \citenamefont {Renders},\ and\ \citenamefont {Saerens}}]{FoussEtAl2007}%
  \BibitemOpen
  \bibfield  {author} {\bibinfo {author} {\bibfnamefont {F.}~\bibnamefont {Fouss}}, \bibinfo {author} {\bibfnamefont {A.}~\bibnamefont {Pirotte}}, \bibinfo {author} {\bibfnamefont {J.-M.}\ \bibnamefont {Renders}},\ and\ \bibinfo {author} {\bibfnamefont {M.}~\bibnamefont {Saerens}},\ }\href@noop {} {\bibfield  {journal} {\bibinfo  {journal} {IEEE Transactions on Knowledge and Data Engineering}\ }\textbf {\bibinfo {volume} {19}},\ \bibinfo {pages} {355} (\bibinfo {year} {2007})}\BibitemShut {NoStop}%
\bibitem [{\citenamefont {Noh}\ and\ \citenamefont {Rieger}(2004)}]{NohRieger2004}%
  \BibitemOpen
  \bibfield  {author} {\bibinfo {author} {\bibfnamefont {J.~D.}\ \bibnamefont {Noh}}\ and\ \bibinfo {author} {\bibfnamefont {H.}~\bibnamefont {Rieger}},\ }\href {https://doi.org/10.1103/PhysRevLett.92.118701} {\bibfield  {journal} {\bibinfo  {journal} {Phys. Rev. Lett.}\ }\textbf {\bibinfo {volume} {92}},\ \bibinfo {pages} {118701} (\bibinfo {year} {2004})}\BibitemShut {NoStop}%
\bibitem [{\citenamefont {Adamic}\ \emph {et~al.}(2001)\citenamefont {Adamic}, \citenamefont {Lukose}, \citenamefont {Puniyani},\ and\ \citenamefont {Huberman}}]{AdamicEtAl2001}%
  \BibitemOpen
  \bibfield  {author} {\bibinfo {author} {\bibfnamefont {L.~A.}\ \bibnamefont {Adamic}}, \bibinfo {author} {\bibfnamefont {R.~M.}\ \bibnamefont {Lukose}}, \bibinfo {author} {\bibfnamefont {A.~R.}\ \bibnamefont {Puniyani}},\ and\ \bibinfo {author} {\bibfnamefont {B.~A.}\ \bibnamefont {Huberman}},\ }\href {https://doi.org/10.1103/PhysRevE.64.046135} {\bibfield  {journal} {\bibinfo  {journal} {Phys. Rev. E}\ }\textbf {\bibinfo {volume} {64}},\ \bibinfo {pages} {046135} (\bibinfo {year} {2001})}\BibitemShut {NoStop}%
\bibitem [{\citenamefont {Capitán}\ \emph {et~al.}(2012)\citenamefont {Capitán}, \citenamefont {Borge-Holthoefer}, \citenamefont {Gómez}, \citenamefont {Martinez-Romo}, \citenamefont {Araujo}, \citenamefont {Cuesta},\ and\ \citenamefont {Arenas}}]{LocalBasedSNavPaper}%
  \BibitemOpen
  \bibfield  {author} {\bibinfo {author} {\bibfnamefont {J.~A.}\ \bibnamefont {Capitán}}, \bibinfo {author} {\bibfnamefont {J.}~\bibnamefont {Borge-Holthoefer}}, \bibinfo {author} {\bibfnamefont {S.}~\bibnamefont {Gómez}}, \bibinfo {author} {\bibfnamefont {J.}~\bibnamefont {Martinez-Romo}}, \bibinfo {author} {\bibfnamefont {L.}~\bibnamefont {Araujo}}, \bibinfo {author} {\bibfnamefont {J.~A.}\ \bibnamefont {Cuesta}},\ and\ \bibinfo {author} {\bibfnamefont {A.}~\bibnamefont {Arenas}},\ }\href {https://doi.org/10.1371/journal.pone.0043694} {\bibfield  {journal} {\bibinfo  {journal} {PLoS ONE}\ }\textbf {\bibinfo {volume} {7}},\ \bibinfo {pages} {e43694} (\bibinfo {year} {2012})}\BibitemShut {NoStop}%
\bibitem [{\citenamefont {Hughes}\ and\ \citenamefont {Ramage}(2007)}]{HughesRamage2007}%
  \BibitemOpen
  \bibfield  {author} {\bibinfo {author} {\bibfnamefont {J.}~\bibnamefont {Hughes}}\ and\ \bibinfo {author} {\bibfnamefont {D.}~\bibnamefont {Ramage}},\ }in\ \href@noop {} {\emph {\bibinfo {booktitle} {Proceedings of the 2007 Joint Conference on Empirical Methods in Natural Language Processing and Computational Natural Language Learning (EMNLP-CoNLL)}}}\ (\bibinfo {year} {2007})\ pp.\ \bibinfo {pages} {776--784}\BibitemShut {NoStop}%
\bibitem [{SM()}]{SM}%
  \BibitemOpen
  \href@noop {} {}\bibinfo {note} {See Supplemental Material at [URL will be inserted by publisher]}\BibitemShut {NoStop}%
\bibitem [{\citenamefont {Banerjee}\ and\ \citenamefont {Pedersen}(2002)}]{banerjee2002lesk}%
  \BibitemOpen
  \bibfield  {author} {\bibinfo {author} {\bibfnamefont {S.}~\bibnamefont {Banerjee}}\ and\ \bibinfo {author} {\bibfnamefont {T.}~\bibnamefont {Pedersen}},\ }in\ \href@noop {} {\emph {\bibinfo {booktitle} {International Conference on Intelligent Text Processing and Computational Linguistics}}}\ (\bibinfo {organization} {Springer},\ \bibinfo {year} {2002})\ pp.\ \bibinfo {pages} {136--145}\BibitemShut {NoStop}%
\bibitem [{\citenamefont {Pesaranghader}\ \emph {et~al.}(2016)\citenamefont {Pesaranghader}, \citenamefont {Matwin}, \citenamefont {Sokolova},\ and\ \citenamefont {Beiko}}]{pesaranghader2016simdef}%
  \BibitemOpen
  \bibfield  {author} {\bibinfo {author} {\bibfnamefont {A.}~\bibnamefont {Pesaranghader}}, \bibinfo {author} {\bibfnamefont {S.}~\bibnamefont {Matwin}}, \bibinfo {author} {\bibfnamefont {M.}~\bibnamefont {Sokolova}},\ and\ \bibinfo {author} {\bibfnamefont {R.~G.}\ \bibnamefont {Beiko}},\ }\href@noop {} {\bibfield  {journal} {\bibinfo  {journal} {Bioinformatics}\ }\textbf {\bibinfo {volume} {32}},\ \bibinfo {pages} {1380} (\bibinfo {year} {2016})}\BibitemShut {NoStop}%
\bibitem [{\citenamefont {Grover}\ and\ \citenamefont {Leskovec}(2016)}]{Grover2016node2vec}%
  \BibitemOpen
  \bibfield  {author} {\bibinfo {author} {\bibfnamefont {A.}~\bibnamefont {Grover}}\ and\ \bibinfo {author} {\bibfnamefont {J.}~\bibnamefont {Leskovec}},\ }in\ \href {https://doi.org/10.1145/2939672.2939754} {\emph {\bibinfo {booktitle} {Proceedings of the 22nd ACM SIGKDD International Conference on Knowledge Discovery and Data Mining}}}\ (\bibinfo  {publisher} {ACM},\ \bibinfo {year} {2016})\ pp.\ \bibinfo {pages} {855--864}\BibitemShut {NoStop}%
\bibitem [{\citenamefont {García-Cuadrillero}(2025)}]{ODCodeRepo}%
  \BibitemOpen
  \bibfield  {author} {\bibinfo {author} {\bibfnamefont {P.}~\bibnamefont {García-Cuadrillero}},\ }\href@noop {} {\bibinfo {title} {Code and data for 'ontological differentiation as a measure of semantic accuracy'}},\ \bibinfo {howpublished} {\url{https://github.com/pablogc1/OD-as-a-measure-of-semantic-accuracy/tree/main}} (\bibinfo {year} {2025})\BibitemShut {NoStop}%
\bibitem [{\citenamefont {Dijkstra}(1959)}]{Dijkstra1959}%
  \BibitemOpen
  \bibfield  {author} {\bibinfo {author} {\bibfnamefont {E.~W.}\ \bibnamefont {Dijkstra}},\ }\href {https://doi.org/10.1007/BF01386390} {\bibfield  {journal} {\bibinfo  {journal} {Numerische Mathematik}\ }\textbf {\bibinfo {volume} {1}},\ \bibinfo {pages} {269} (\bibinfo {year} {1959})}\BibitemShut {NoStop}%
\end{thebibliography}%

\bibliographystyle{apsrev4-2} 
\end{document}